\documentclass[times]{aastex631}
\usepackage{physics}
\usepackage{mathrsfs}
\usepackage{lipsum}
\let\tablenum\relax
\usepackage{siunitx}

\shortauthors{Songsheng et al.}

\graphicspath{{./}{figures/}}

\begin{document}

\title{Differential Interferometric Signatures of Close Binaries of Supermassive Black Holes in Active Galactic Nuclei: \uppercase\expandafter{\romannumeral2}. Merged Broad Line Regions}

%\correspondingauthor{August Muench}
%\email{greg.schwarz@aas.org, gus.muench@aas.org}

\author[0000-0003-4042-7191]{Yu-Yang Songsheng}
\affiliation{Key Laboratory for Particle Astrophysics, Institute of High Energy Physics, Chinese Academy of Sciences, 19B Yuquan Road, Beijing 100049, People's Republic of China}
\affiliation{Dongguan Neutron Science Center, 1 Zhongziyuan Road, Dongguan 523808, People's Republic of China}

\author[0000-0001-9449-9268]{Jian-Min Wang}
\affiliation{Key Laboratory for Particle Astrophysics, Institute of High Energy Physics, Chinese Academy of Sciences, 19B Yuquan Road, Beijing 100049, People's Republic of China}
\affiliation{University of Chinese Academy of Sciences, 19A Yuquan Road, Beijing 100049, People's Republic of China}
\affil{National Astronomical Observatories of China, Chinese Academy of Sciences, 20A Datun Road, Beijing 100020, China}

%\author{...}

\begin{abstract}
Pairs of supermassive black holes (SMBHs) at different stages are natural results of galaxy mergers in the hierarchical framework of galaxy formation and evolution.
However, identifications of close binaries of SMBHs (CB-SMBHs) with sub-parsec separations in observations are still elusive.
Recently, unprecedented spatial resolutions achieved by GRAVITY/GRAVITY+ onboard The Very Large Telescope Interferometer through spectroastrometry (SA) provide new opportunities to resolve CB-SMBHs.
Differential phase curves of CB-SMBHs with two independent broad-line regions (BLRs) are found to have distinguished characteristic structures from a single BLR \citep{songsheng2019}.
Once the CB-SMBH evolves to the stage where BLRs merge to form a circumbinary BLR, it will hopefully be resolved by the pulsar timing array (PTA) in the near future as sources of nano-hertz gravitational waves.
In this work, we use a parameterized model for circumbinary BLRs to calculate line profiles and differential phase curves for SA observations.
We show that both profiles and phase curves exhibit asymmetries caused by the Doppler boosting effect of accretion disks around individual black holes, depending on the orbital parameters of the binary and geometries of the BLR.
We also generate mock SA data using the model and then recover orbital parameters by fitting the mock data.
Degeneracies between parameters contribute greatly to uncertainties of parameters but can be eased through joint analysis of multiple-epoch SA observations and reverberation mappings.
\end{abstract}



\section{Introduction} \label{sec:intro}
The existence of supermassive black holes (SMBHs) at the center of nearly every galaxy has been confirmed by both star/gas dynamics of local galaxies \citep{genzel2010,kormendy2013} and reverberation mapping (RM) of distant active galaxies \citep[e.g.][]{kaspi2000,du2016}.
Galaxies grow through frequent mergers according to the hierarchical model of galaxy formation and evolution \citep{white1978,lacey1993,patton2002,lin2004,conselice2014}, leading to the formation of binary SMBHs.
Dual AGNs with separations ranging from several parsecs to several kiloparsecs have been identified through direct imaging \citep[e.g.][]{rodriguez2006,comerford2009}.
However, close binaries of SMBHs (CB-SMBHs) with sub-parsec separations have not been identified conclusively.
The evolution mechanism of SMBH binaries after its separation decreases below a few parsecs is still debated \citep[e.g.][]{milosavljevic2003,escala2005,cuadra2009}, and the existence and distribution of CB-SMBHs in the universe remain uncertain.
Identification of CB-SMBHs would be a litmus test for the theory of galaxy mergers, resolving the long-standing ``final parsec problem'' completely.
Additionally, CB-SMBHs are dominating sources of nano-hertz gravitational waves (GWs) background in our universe, which can be detected by pulsar timing arrays (PTAs) \citep{sazhin1978,jenet2006,antoniadis2022}.
Sufficiently strong gravitational waves produced by massive CB-SMBHs can even be resolved individually \citep{sesana2009}.
If the GW source can also be identified by electromagnetic characteristics in advance, we can not only improve the sensitivity of the PTA through target search \citep{arzoumanian2020,songsheng2021}, but also restrict orbital parameters of the binary more precisely to test the physics of nano-hertz GWs through joint analysis of multimessenger signals.

Several electromagnetic characteristics have been proposed to search CB-SMBHs indirectly, among which the most widely used are periodic variations of the flux.
In CB-SMBHs, periodic accretion of the gas in circumbinary disks or the Doppler boosting effect of the mini-disks around individual black holes \citep{dorazio2015} can modulate the radiation flux periodically.
Hundreds of candidates of CB-SMBHs with periodic variations have been selected from several photometric surveys, such as Catalina Real-time Transient Survey \citep{graham2015}, Palomar Transient Factory \citep{charisi2016}, Pan-STARRS1 Medium Deep Survey \citep{liu2019} and Zwicky Transient Facility \citep{chen2022}.
However, stochastic variabilities of quasars driven by thermal fluctuations \citep{kelly2009,kelly2011} or other random processes can mimic periodic signals in a finite time span \citep{vaughan2016,goyal2018}.
Generally, the light curve should span at least three times the claimed period to avoid false periodicities \citep{li2018}.
Other approaches mainly rely on the double-peaked or asymmetric profiles of broad emission lines \citep{shen2010,popovic2012,nguyen2016}, which reflect the complex dynamics of the gas in the gravitational fields of the binary black holes.
However, the broad-line region (BLR) around a single black hole can also be complicated enough to produce double-peaked or asymmetric profiles \citep{eracleous1994, wang2017}, and more information is needed to break the degeneracy.
For example, long-term spectroscopic monitoring of NGC 4151 \citep{bon2012}, NGC 5548 \citep{li2016}, and Ark 120 \citep{li2019} show that their line profiles vary periodically, making them promising candidates for CB-SMBHs.
Particularly, binary candidates with relatively extreme mass ratios and at separations between $0.1$ and $\SI{1}{pc}$ can be selected from large populations of AGNs by detecting the time-dependent velocity offsets through multi-epoch spectroscopy\citep{kelley2021}.
RM is also an efficient way to probe the geometry and velocity field of BLR through 2D transfer functions, which reflects the response lags of clouds with different line-of-sight (LOS) velocities to continuum variation \citep{blandford1982}.
2D transfer functions of BLRs in binary black holes are distinguished from those around single black holes since the broad emission line will respond to the continuum variation asymmetrically or out-of-sync at the red and blue wing \citep{wang2018,songsheng2020,kovacevic2020a,ji2021}.

The difficulty to identify CB-SMBHs lies in their extremely small angular sizes, which can not be resolved through direct imaging.
Recently, the GRAVITY onboard The Very Large Telescope Interferometer (VLTI) has achieved an astrometric accuracy of $\sim \SI{10}{\mu as}$ \citep{gravity2017} and spatially resolved BLRs of several AGNs using spectroastrometry (SA) \citep{gravity2018,gravity2020,gravity2021}.
By measuring the variation of the interferometric phase with wavelength across the broad emission line, SA detects the angular displacements of photocenters of BLR clouds with different LOS velocities, which further reflects the geometry and dynamics of the BLR.
The differential phase curve of a single BLR around a SMBH usually shows a symmetric ``S-like'' shape with a peak and valley.  
When it is applied to CB-SMBHs with separated BLRs, the binary BLRs can be partially resolved through phase plateaus or extra peaks between the major peak and valley in interferometric phase curves \citep{songsheng2019,kovacevic2020b}.
For SMBHs with masses of $\num{e8} - \num{e9}M_{\sun}$, the size of their BLRs are usually larger than $\SI{0.1}{pc}$ \citep{bentz2013}.
If BLRs of the two black holes are still detached, frequencies of GWs emitted by them are too low to be detected using PTA.
For multimessenger observation of CB-SMBHs, we must involve those with merged BLRs.
On the one hand, CB-SMBH in a circumbinary disk can be diagnosed through their complex emission-line profiles generated by two mini-disks that are gravitationally bound to the individual black holes and the circumbinary disk \citep{nguyen2016,nguyen2019,nguyen2020}.
On the other hand, the orbital velocities of these CB-SMBHs usually reach a few percent of the speed of light, and the relativistic Doppler boosting effect of the mini-disks around individual black holes will be observably \citep{charisi2022}.
As a result, the circumbinary BLR will be illuminated anisotropically, and the distribution of emissivities of clouds will also be asymmetric.
The RM of circumbinary BLR has been studied thoroughly in \citet{ji2021}, showing that the broad emission lines' blue and red wings are enhanced and weakened periodically but with a phase difference.
Due to the coupling between timescales of light propagation and orbital motion, the RM cannot be fully described by a transfer function, making the identification not straightforward.
Fortunately, the SA probes the dynamics of the BLR by taking a snapshot and so is free of the complications in the time domain, providing a different perspective to recognize CB-SMBHs.

In this paper, we established a parametric model for circumbinary BLRs and studied the dependence of differential phase curves on model parameters.
We also fit the model to the simulated interferometric data and recover the orbital parameters of the CB-SMBHs.
The paper is scheduled as the following. 
We present the fundamental formulations of the circumbinary BLR model in \S2. 
The differential phase curves under different model parameters and the recovery of model parameters using mock data are provided in \S3.
Discussions on the issues of binary black holes are given in \S4. 
We conclude in the last section. 

\section{Methodology} \label{sec:floats}
\subsection{Circumbinary BLR}
The geometry and kinematics of the BLR around a single SMBH have been thoroughly studied through RM in the last three decades.
The velocity-resolved delay map constructed by maximum entropy method \citep[MEM;][]{horne1994, xiao2018} or pixon-based method \citep{li2021} can be used to identify the structure of the BLR, and demonstrate that flattened disk with Keplerian rotation is common in Seyfert galaxies \citep{grier2013, du2016}.
Multiple RMs of a few AGNs \citep[such as in NGC 5548, 3C 390.3, and NGC 7469;][]{lu2016,wandel1999} also show evidence for virialized BLRs.
If the separation between the binary black hole is much larger than the sizes of both BLRs, the BLR of each black hole can be described by a detached flattened disk and the velocity of the cloud will be the superposition of the individual virial motion and orbital motion of the binary system \citep{shen2010,popovic2012,wang2018,songsheng2019,songsheng2020}.

As the binary of black holes evolves closer, two BLRs will merge to form a common circumbinary BLR and the dynamics of clouds will be much more complicated in the gravitational field of the binary.
The geometry and kinematics of the BLR could be obtained by integrating the equation of motion of the circular restricted three-body problem until a quasi-equilibrium
is established \citep{shen2010}.
However, the numeric integration is time-consuming and orbits of clouds may be chaotic and even become unbound in the long run.

\begin{figure}
    \centering
    \includegraphics[width=0.9\linewidth]{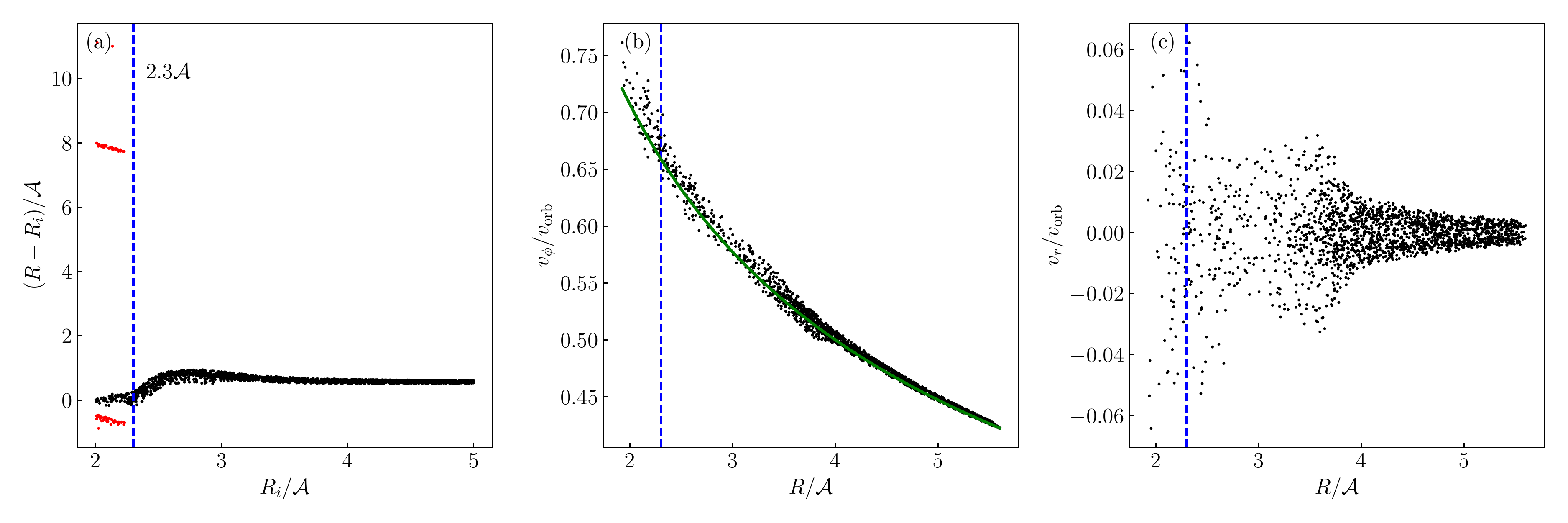}
    \caption{Spatial and kinematic distribution of clouds in BLR after dynamical evolution.
    (a) The $x$-axis is the initial radial positions of the clouds (in units of the binary separation $\mathcal{A}$.
    The $y$-axis is the deviation of the final positions relative to the initial positions.
    Black dots represent clouds that are still in the circumbinary BLR finally, while red dots are those either unbounded or captured by black holes.
    The blue dashed line marks $R = 2.3\mathcal{A}$. 
    (b) The $x$-axis is the final radial position of the clouds that are still in the BLR.
    The $y$-axis is the final rotational velocities of the clouds (in units of the orbital velocity of the binary $v_{\rm orb}$).
    The blue thick line represents the Keplerian velocities at different radial positions.
    (c) The $x$-axis and $y$-axis are the final radial positions and velocities of the clouds respectively.}
    \label{fig:model}
\end{figure}

The long-term stability of test particles in the gravitational field of a binary system has been examined both analytically \citep{szebehely1980} and numerically \citep{holman1999} for various mass ratios and binary eccentricities ($e$).
For the binary with $e=0$, test particles in circumbinary orbits will remain stable after $10^4$ binary periods if the orbital radius is larger than $2.3$ times the binary semimajor axis.
So we assume the circumbinary BLR as a flattened disk with an inner radius larger than $2.3\mathcal{A}$, where $\mathcal{A}$ is the binary semimajor axis.
The rotational velocity of the cloud is simply given by $V = \sqrt{GM_{\bullet}/R}$, where $G$ is the gravitational constant, $R$ is the distance of the cloud to the center of mass of the binary, and $M_{\bullet}$ is the total mass of the binary.
To check the self-consistency of our simplified BLR model, we let the system evolves dynamically under Newtonian gravity \citep{murray1999}.
Initially, $2000$ clouds distribute uniformly from $R = 2\mathcal{A}$ to $R = 5\mathcal{A}$.
After integrating the system dynamically for $1000$ binary periods using Runge-Kutta-Fehlberg method, we obtain the result shown in Fig. \ref{fig:model}.
In Fig. \ref{fig:model}(a), all clouds with initial distances larger than $2.3\mathcal{A}$ are still bound after evolution but move a little bit outward collectively, while some clouds within $2.3\mathcal{A}$ either become unbound or accreted by the black hole.
In Fig. \ref{fig:model}(b) and (c), rotational velocities of survival clouds still follow Keplerian law tightly and their radial velocities are small compared to the orbital velocity of the binary.
All clouds are limited in the orbital plane of the binary and their velocities also have no component along the direction perpendicular to the orbital plane in the test, but the model is also valid as long as the opening angle of the BLR is small.
We neglect the interaction between clouds, but our model is generally consistent with the model of a thin circumbinary gaseous disk governed by gas pressure and viscosity \citep{artymowicz1994}, except that the mass flow through the gap is neglected here \citep{artymowicz1996}.

Finally, we assume the radial emissivity distribution of clouds in BLR is generated by a shifted $\Gamma$-distribution \citep{pancoast2014},
\begin{equation}
    R = 2.3\mathcal{A} + \mathscr{F} (R_{\rm BLR} - 2.3\mathcal{A}) + \Gamma_0\beta^2(1-F)(R_{\rm BLR} - 2.3\mathcal{A}),
\end{equation}
where $R$ is the distance of the cloud to the center of mass of the binary, $R_{\rm BLR}$ is the mean radius, $\mathscr{F}=(R_{\rm in} - 2.3\mathcal{A}) / (R_{\rm BLR} - 2.3\mathcal{A})$ is the modified fraction of the inner to the mean radius, $R_{\rm in}$ is the inner radius, $\beta$ is the shape parameter, and $\Gamma_0 = p(x|\beta^{-2},1)$ is a random number drawn from a $\Gamma$-distribution
\begin{equation}
    p(x|\alpha,x_0) = \frac{x^{\alpha-1}\exp(-x/x_0)}{\Gamma(\alpha)x_0^{\alpha}}
\end{equation}
with $\alpha = \beta^{-2}$ and $x_0 = 1$.
% Note that $R_{\rm in} = FR_{\rm BLR}$ should be larger than $2.3a$ in our model.

For a typical local massive CB-SMBH, we assume its angular size distance, total mass and separation are $D_{\rm A} = \SI{200}{Mpc}$, $M_{\bullet} = \num{e9} M_{\sun}$ and $\mathcal{A} = \SI{0.02}{pc}$ respectively.
On the one hand, the angular separation of the binary is
\begin{equation}
    \xi_{\rm a} = \num{21} \left(\frac{\mathcal{A}}{\SI{0.02}{pc}}\right) \left(\frac{D_{\rm A}}{\SI{200}{Mpc}}\right)^{-1} \si{\mu as},
\end{equation}
which can be resolved by GRAVITY/GRAVITY+ through SA.
On the other hand, the frequency and strain amplitude of GWs emitted by the binary is
\begin{equation}
    f_{\rm GW} = \num{7.6} (1+z)^{-1} \left(\frac{M_{\bullet}}{\num{e9} M_{\sun}}\right)^{1/2} \left(\frac{\mathcal{A}}{\SI{0.02}{pc}}\right)^{-3/2} \si{nHz}
\end{equation}
and
\begin{equation}
    h_{\rm c} = \num{1.1e-15} (1+z)^{-1} q (1+q)^{-2} \left(\frac{M_{\bullet}}{\num{e9} M_{\sun}}\right)^2 \left(\frac{\mathcal{A}}{\SI{0.02}{pc}}\right)^{-1} \left(\frac{D_{\rm A}}{\SI{200}{Mpc}}\right)^{-1},
\end{equation}
where $z$ is the cosmological redshift of the binary, $q \equiv M_1 / M_2$ is the mass ratio between the primary ($M_1$) and secondary ($M_2$) black hole.
So GWs emitted by the binary are expected to be detected by PTA in the era of SKA\citep{moore2015}.

\subsection{Doppler boosting effect}
For our typical CB-SMBH, the ratio between the orbital velocity and speed of light is given by
\begin{equation}
    \beta_{\rm a} = 0.05 \left(\frac{M_{\bullet}}{\num{e9} M_{\sun}}\right)^{1/2} \left(\frac{\mathcal{A}}{\SI{0.02}{pc}}\right)^{-1/2}.
\end{equation}
As the orbital velocity of the binary reaches a few percent of the light speed, the Doppler boosting will make the accretion disk a non-isotropic source of ionization radiation and the surface brightness of the BLR will also become asymmetric.
We quantify this effect in a similar framework as that in \cite{ji2021}.

\begin{figure}
    \centering
    \includegraphics[width=0.5\linewidth]{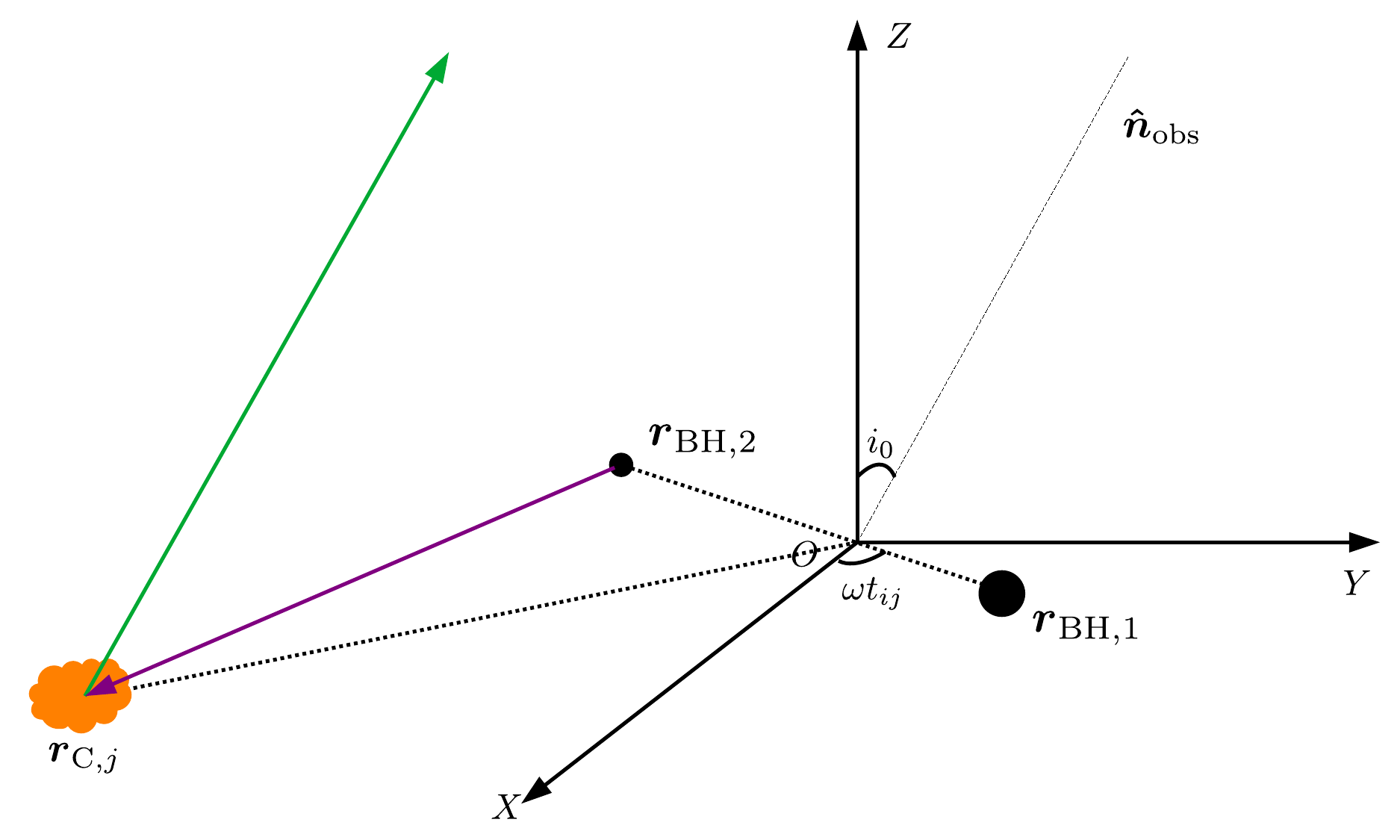}
    \caption{Coordinate system of the circumbinary BLR model. 
    The origin $O$ locate at the mass center of the binary, and $O-XY$ is the orbital plane of the binary. 
    The direction of line-of-sight is in the $O-YZ$ plane and its angle with the $OZ$ axis is $i_0$. At the time $t_{ij}$, the accretion disk around the secondary black hole at $\vb*{r}_{\rm BH,2}$ emits an ultraviolet photon, whose trace is represented by the purple arrow}.
    The photon then reaches the BLR cloud at $\vb*{r}_{\mathrm{C},j}$ and excites the hydrogen atom.
    The recombination line emission propagating along the line of sight (the green arrow) finally reaches the distant observer. \label{fig:coordinates}
\end{figure}

The coordinate system is shown in Fig. \ref{fig:coordinates}.
At time $t$, coordinates of the primary and secondary black hole are $\vb*{r}_{\rm BH,1} = (1+q)^{-1}\mathcal{A}(\cos\omega t,\sin\omega t,0)$ and $\vb*{r}_{\rm BH,2} = -q(1+q)^{-1} \mathcal{A} (\cos\omega t,\sin\omega t,0)$ respectively, where $\omega=(GM_{\bullet})^{1/2}\mathcal{A}^{-3/2}$ is the angular velocity of the binary.
The corresponding velocities of the two black holes are $\vb*{v}_{\rm BH,1} = (1+q)^{-1}\omega \mathcal{A}(-\sin\omega t,\cos\omega t,0)$ and $\vb*{v}_{\rm BH,2} = -q(1+q)^{-1} \omega \mathcal{A} (-\sin\omega t,\cos\omega t,0)$.

If the accretion disk of the $i$-th black hole emits an ultraviolet photon at time $t_{ij}$, it will reach the $j$-th cloud at $\vb*{r}_{{\rm C},j}$ with velocity $\vb*{v}_{{\rm C},j}$ at time $t_{ij} + \abs{\vb*{r}_{{\rm C},j} - \vb*{r}_{{\rm BH},i}} / c$ and excite the hydrogen atom there.
The frequency of the photon will be boosted by a factor
\begin{equation}
    D_{ij} = \frac{1}{\gamma_{ij}(1+\beta_{\parallel ij})},
\end{equation}
where
\begin{equation}\label{eq:vel_para}
    \beta_{\parallel ij} = \frac{(\vb*{v}_{{\rm C},j} - \vb*{v}_{{\rm BH},i})\vdot(\vb*{r}_{{\rm C},j} - \vb*{r}_{{\rm BH},i})}{\abs{\vb*{r}_{{\rm C},j} - \vb*{r}_{{\rm BH},i}}c}
\end{equation}
is the projection of the $j$-th cloud's relative velocity along the direction of its relative position to the $i$-th black hole
and
\begin{equation}
    \gamma_{ij} = \frac{1}{\sqrt{1 - \abs{\vb*{v}_{{\rm C},j} - \vb*{v}_{{\rm BH},i}}/c^2}}
\end{equation}
is the corresponding Lorentz factor.
According to the conservation of $I_{\nu}/\nu^3$ for the propagation of light in a vacuum, where $\nu$ is the frequency of the photon and $I_{\nu}$ is the specific intensity at $\nu$, the ionizing flux received by the $j$-th cloud is then given by
\begin{equation}
    F_{{\rm C},ij}(\nu) \propto F_{{\rm AD}_i}(\nu D_{ij}^{-1}) D_{ij}^{3} \frac{\abs{\vb*{r}_{{\rm C},j}}^2}{\abs{\vb*{r}_{{\rm C},j} - \vb*{r}_{{\rm BH},i}}^2} = F_{{\rm AD}_i}(\nu) D_{ij}^{3-\alpha_{\nu,i}} \frac{\abs{\vb*{r}_{{\rm C},j}}^2}{\abs{\vb*{r}_{{\rm C},j} - \vb*{r}_{{\rm BH},i}}^2},
\end{equation}
where $F_{{\rm AD}_i}(\nu) \propto \nu^{\alpha_{\nu, i}}$ is the intrinsic SED of the accretion disk of the $i$-th black hole, and the term ${\abs{\vb*{r}_{{\rm C},j}}^2}/{\abs{\vb*{r}_{{\rm C},j} - \vb*{r}_{{\rm BH}, i}}^2}$ accounts of the variation of the opening angle spanned by the $j$-th cloud due to the motion of the $i$-th black hole.
Note that we assume the intrinsic variation of $F_{\nu,{\rm AD}_i}$ is negligible compared to the variation of the Doppler boosting factor.

The photon of the recombination line reaches the observer at a time 
\begin{equation}
    T_{\rm obs} = T_{ij}(t_{ij}) \equiv (1+z)\left[t_{ij} + \frac{\abs{\vb*{r}_{{\rm C},j} - \vb*{r}_{{\rm BH},i}}}{c} - \frac{\vb*{r}_{{\rm C},j}\vdot \vb*{n}_{\rm obs}}{c}\right] + T_0,
\end{equation}
where $\vb*{n}_{\rm obs}$ is the unit vector pointing from the AGN to the distant observer, and $T_0$ is an overall delay for a photon propagating from the center of mass of the binary to the observer.
The Doppler boosting factor due to the relative motion of the cloud relative to the observer is given by
\begin{equation}
    D_j = \frac{1}{\gamma_{j}(1+\beta_{\parallel j})},
\end{equation}
where $\beta_{\parallel j} = -\vb*{v}_{{\rm C},j} \vdot \vb*{n}_{\rm obs}/c$ and $\gamma_{j} = (1-\abs{\vb*{v}_{{\rm C},j}}^2/c^2)^{-1/2}$.
So the frequency of the photon received by the observer is $\nu_{\rm c}D_j(1+z)^{-1}$, where $\nu_{\rm c}$ is the frequency of the emission line in the rest frame.
The corresponding flux is
\begin{equation}\label{eq:clouds_flux}
    F_{ij} \propto F_{{\rm AD}_i} D_{ij}^{3-\alpha_{\nu,i}} D_j^3 \frac{\abs{\vb*{r}_{{\rm C},j}}^2}{\abs{\vb*{r}_{{\rm C},j} - \vb*{r}_{{\rm BH},i}}^2}.
\end{equation}
Eq. (\ref{eq:clouds_flux}) describes relative emissivities of clouds at different positions, which mainly depends on the Doppler factor $D_{ij}$.
In the direction where the accretion disk moves toward the cloud, the relative emissivity will boost significantly.

\subsection{Spectroastrometry}
Suppose the optical interferometer takes the observation at time $T_{\rm obs}$.
For each black hole and cloud, we should solve the equation $T_{ij}(t_{ij}) = T_{\rm obs}$ for $t_{ij}$. The equation can be further reorganized as
\begin{equation}\label{eq:aber}
    \phi_{ij} = \phi_0 - \phi_j - \beta_a \frac{\abs{\vb*{r}_{{\rm C},j} - \vb*{r}_{{\rm BH},i}(\phi_{ij})} - \abs{\vb*{r}_{{\rm C},i}}}{\mathcal{A}}
\end{equation}
where $\phi_{ij} \equiv \omega t_{ij}$, $\phi_0 \equiv \omega(T_{\rm obs} - T_0)/(1+z)$, $\phi_j \equiv \omega(\abs{\vb*{r}_{{\rm C},j}} - \vb*{r}_{{\rm C},j}\vdot \vb*{n}_{\rm obs})/c$ and $\beta_a = (GM_{\bullet})^{1/2}\mathcal{A}^{-1/2}c^{-1}$.
Note that $\beta \sim 0.1$ and $\abs{\abs{\vb*{r}_{{\rm C},j} - \vb*{r}_{{\rm BH},i}} - \abs{\vb*{r}_{{\rm C},i}}} < |\vb*{r}_{{\rm BH},i}| < \mathcal{A}$. The last term of the Eq.(\ref{eq:aber}) is a small quantity and so the equation can be solved iteratively as
\begin{equation}
    \phi_{ij}^{(0)} = \phi_0 - \phi_j \qc \phi_{ij}^{(k+1)} = \phi_0 - \phi_j - \beta_a \frac{\abs{\vb*{r}_{{\rm C},j} - \vb*{r}_{{\rm BH},i}(\phi_{ij}^{(k)})} - \abs{\vb*{r}_{{\rm C},i}}}{\mathcal{A}}
\end{equation}
until $\abs*{\phi_{ij}^{(k+1)} - \phi_{ij}^{(k)}} < \num{e-3}$.
Once all $\phi_{ij}$ are obtained, the wavelength-dependent photocenter of the BLR can be figured out as
\begin{equation}
    \vb*{\epsilon}_{\ell}(\lambda) = \frac{\sum_{i=1}^{2}\sum_{j}F_{ij}\delta(\lambda -\lambda_j)\vb*{\theta}_j}{\sum_{i=1}^{2}\sum_{j}F_{ij}\delta(\lambda -\lambda_j)},
\end{equation}
where $\lambda_j = (1+z)\lambda_{\rm c}D_j^{-1}$ is the wavelength of the photon emitted by the $j$-th cloud, $\lambda_{\rm c}$ is the wavelength of the emission line at rest frame; $\vb*{\theta}_j = [\vb*{r}_{{\rm C},j} - (\vb*{r}_{{\rm C},j}\vdot\vb*{n}_{\rm obs})\vb*{n}_{\rm obs}]/D_{\rm A}$ is the angular displacement of the cloud in the tangent plane of the sky.

The continuum emission in the K-band, which is covered by GRAVITY, mainly originates from the hot dust near the sublimation radius \citep{kishimoto2009}, which is systematically larger than the size of the BLR by about a factor of four to five from infrared RM and interferometry \citep{koshida2014,gravity2020}.
Since the size of the dust torus is much larger than the separation between the binary, the photocenter of the continuum $\vb*{\epsilon}_{\rm c}$ will keep constant as the binary rotates.
$\vb*{\epsilon}_{\rm c}$ is non-dynamic and independent of wavelength, only causing vertical shifts of the interferometric phase curve.
To highlight the characteristics of the differential phase curve of the circumbinary BLR, we assume the location of the continuum photocenter to be the mass center of the binary, i.e. $\vb*{\epsilon}_{\rm c} = \vb*{0}$.
The differential phase curve will be
\begin{equation}\label{eq:phase_curve}
    \phi_{\rm diff}(\lambda) = -2\pi f_{\ell}(\lambda) \frac{\vb*{\epsilon}_{\ell}(\lambda)\vdot\vb*{B}}{\lambda},
\end{equation}
where $f_{\ell} = F_{\ell} / (F_{\ell} + F_{\rm c})$ is the ratio of the emission line flux $F_{\ell}$ to the total flux $F_{\ell} + F_{\rm c}$ in each wavelength channel and $\vb*{B}$ is the sky-projected baseline of the interferometer \citep{gravity2018}.

All independent parameters used in the circumbinary BLR model are shown in Table \ref{tab:CB-BLR}.
For convenience, we define a dimensionless BLR size $r_{\rm BLR} \equiv R_{\rm BLR} / \mathcal{A}$ in Table \ref{tab:CB-BLR}.
Once the angular size distance of the AGN is fixed, other model parameters, such as binary separation, binary period, and total mass, can be calculated directly.
% \begin{equation}\label{eq:binary_separation}
%     \mathcal{A} = \num{1.9e-2} \left(\frac{\xi_{\rm a}}{\SI{20}{\mu as}}\right) \left(\frac{D_{\rm A}}{\SI{200}{Mpc}}\right) \si{pc}.
% \end{equation}
% \begin{equation}\label{eq:binary_period}
%     P = \num{7.9} \left(\frac{\xi_{\rm a}}{\SI{20}{\mu as}}\right) \left(\frac{D_{\rm A}}{\SI{200}{Mpc}}\right)  \left(\frac{\beta}{0.05}\right)^{-1} \si{yr}.
% \end{equation}
% \begin{equation}\label{eq:binary_mass}
%     M_{\bullet} = \num{e9} \left(\frac{\xi_{\rm a}}{\SI{20}{\mu as}}\right) \left(\frac{D_{\rm A}}{\SI{200}{Mpc}}\right)  \left(\frac{\beta}{0.05}\right)^{2} M_{\sun}.
% \end{equation}

\begin{deluxetable}{llllll}
    \tablecaption{Independent parameters used in the circumbinary BLR model \label{tab:CB-BLR}}
    \tablewidth{0pt}
    \tablehead{\colhead{Parameters} & \colhead{Units}  & \colhead{Meanings} & \colhead{Fiducial values} & \colhead{Prior ranges} & \colhead{Prior distribution}}
    \startdata
    $\beta_{\rm a}$ & light speed & orbital speed of the binary &$0.05$ & $[\num{e-3}, 1]$ & log uniform \\
    $\xi_{\rm a}$ & $\si{\mu as}$ & angular size of the binary separation & $20$ & $[1,\num{e3}]$ & log uniform \\
    $\phi_0$ & $\si{\degree}$ & orbital phase of the binary & $90$ & $[0, 360]$ & uniform \\
    $q$ & & mass ratio (primary to secondary) & $2.33$ & $[1, 10]$ & uniform \\
    $f$ & & luminosity ratio (primary to secondary) & $0$ & $[0, 1]$ & uniform \\
    $r_{\rm BLR}$ & & mean size of the BLR in units of $\mathcal{A}$ & $4$ & $[2.3,\num{e3}]$ & log uniform \\
    $\mathscr{F}$ & & modified fractional inner radius & $0.1$ & $[0,1]$ & uniform \\
    $\beta$ & & radial distribution shape parameter & $1.0$ & $[0,4]$ & uniform \\
    $i$ & $\si{\degree}$ & inclination angle of the LOS & $30$ & $[0,90]$ & $\cos i$ uniform\\
    $\theta_{\rm opn}$ & $\si{\degree}$ & half opening angle of the BLR & $10$ & $[0,90]$ & uniform \\
    $\rm PA$ & $\si{\degree}$ & position angle of the projected baseline & $0$ & $[0, 360]$ & uniform\\
    $D_{\rm A}$ & $\si{Mpc}$ & angular size distance of the AGN & $200$ & & fixed  \\
    $\alpha_{\nu,i}$ & & spectral index of the intrinsic SED & $-2$ & & fixed  \\
    \enddata
\end{deluxetable}

\section{Results}
\subsection{Dependence on model parameters}
\begin{figure}
    \centering
    \includegraphics[width=\textwidth]{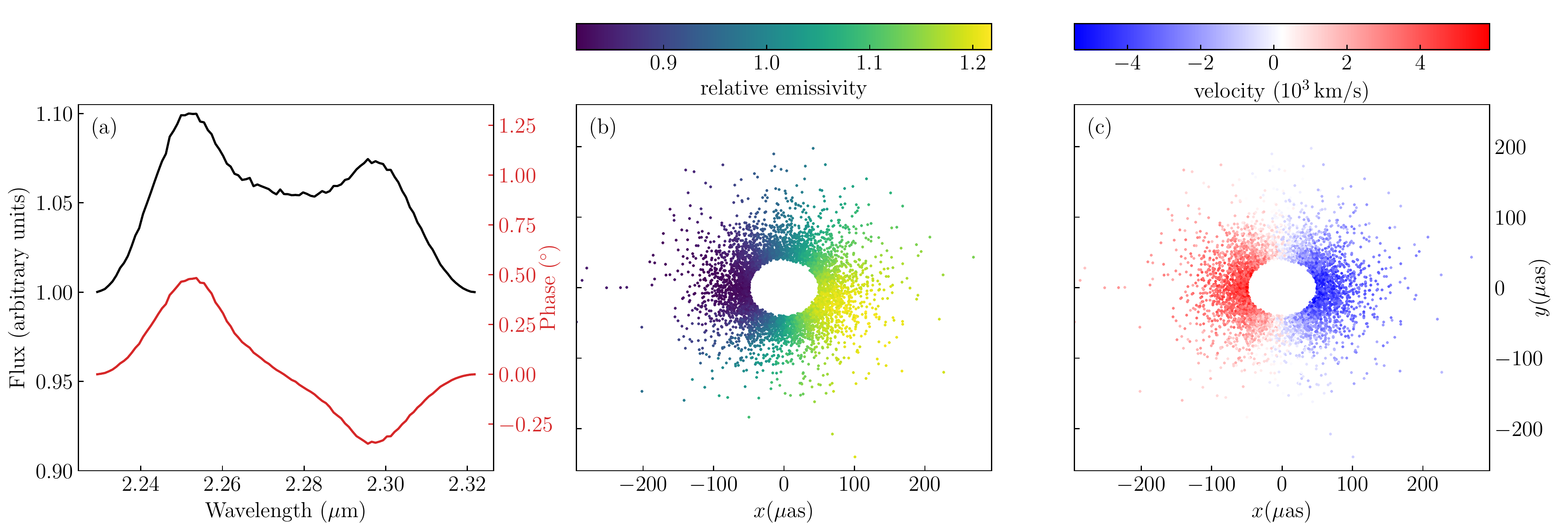}
    \caption{(a) Line profile and differential phase curve of the fiducial BLR model. 
    (b) Distribution of relative emissivities (Eq. \ref{eq:clouds_flux}) of BLR clouds when projected onto the plane tangent to the LOS.
    (c) Distribution of LOS velocities of BLR clouds when projected onto the plane tangent to the LOS.
    \label{fig:fiducial}}
\end{figure}

Firstly, we use the fiducial values in Table \ref{tab:CB-BLR} as parameters of the merged BLR model to calculate the profile and differential phase curve.
As shown in Fig. \ref{fig:fiducial} (a), both the profile and the phase curve become asymmetric due to the Doppler boosting effect.
Distributions of relative intensities and line-of-sight velocities of BLR clouds projected onto the celestial sphere are presented in Fig. \ref{fig:fiducial} (b) and (c) respectively.
The blue-shifted clouds are boosted by about $\SI{10}{\percent}$ while the red-shifted clouds are diminished, so the blue-shifted peak in the profile is higher than the red-shifted peak.
The velocity gradient field of clouds is still symmetric, but the asymmetric $f_{\ell}(\lambda)$ makes the blue-shifted peak in the phase curve high and narrow and the red-shifted valley shallow and wide.

\begin{figure}
    \centering
    \includegraphics[width=\textwidth]{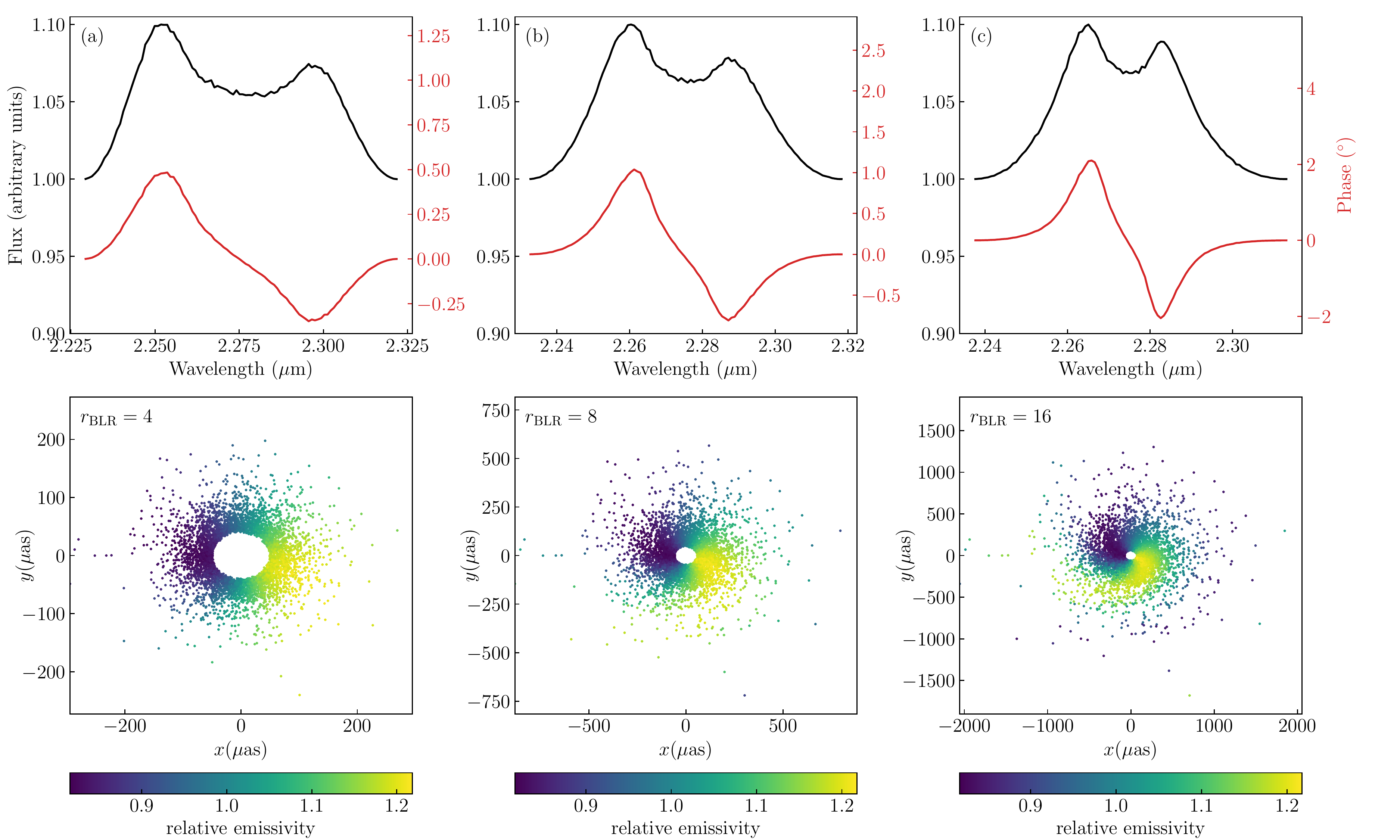}
    \caption{Dependence of line profile and differential phase curve on the BLR size. \label{fig:para_r}}
\end{figure}

Now we increase the mean radius of the BLR from $4a$ to $8a$ and $16a$, and the result is shown in Fig. \ref{fig:para_r}.
As the BLR gets larger, both the profile and phase curve become less asymmetric.
When the size of the BLR is much larger than the separation between the binary, velocities of BLR clouds will be much smaller than that of the binary, and the equation \ref{eq:vel_para} and \ref{eq:aber} can be approximated by
\begin{equation}
    \beta_{\parallel ij} \approx -\frac{\vb*{v}_{{\rm BH},i}(\phi_{ij}) \vdot \vb*{r}_{{\rm C},j}}{\abs{\vb*{r}_{{\rm C},j}}c}  \qc \phi_{ij} \approx \phi_0 - \phi_j = \phi_0 - \beta_a (\abs{\vb*{r}_{{\rm C},j}} - \vb*{r}_{{\rm C},j}\vdot \vb*{n}_{\rm obs})/\mathcal{A}.
\end{equation}
To further simplify the equation, we consider only the clouds in the equatorial plane and suppose the inclination of the LOS is small.
For the secondary black hole, we have
\begin{equation}
    \beta_{\parallel 2j} \approx -\mu \beta_a \sin(\phi_{2j} - \varphi_{{\rm C},j}) \qc \phi_{2j} = \phi_0 - \beta_a \frac{\abs{\vb*{r}_{{\rm C},j}}}{\mathcal{A}}
\end{equation}
where $\varphi_{{\rm C},j}$ is the azimuthal angle of the cloud in the equatorial plane.
For clouds having the same $\varphi_{{\rm C},j} + \beta_a \abs{\vb*{r}_{{\rm C},j}} / \mathcal{A}$, their Doppler boosting factors will be the same, so a spiral structure will develop in the emissivity distribution of clouds, as shown in the second row of Fig. \ref{fig:para_r}.
We can also infer that the larger the rotational velocity of the binary is, the tighter the spiral arm is wound.
As the spiral arm appears in the emissivity distribution, clouds with similar LOS velocities will have a wide range of emissivities, smearing the asymmetry in the profile and phase curve.

\begin{figure}
    \centering
    \includegraphics[width=\textwidth]{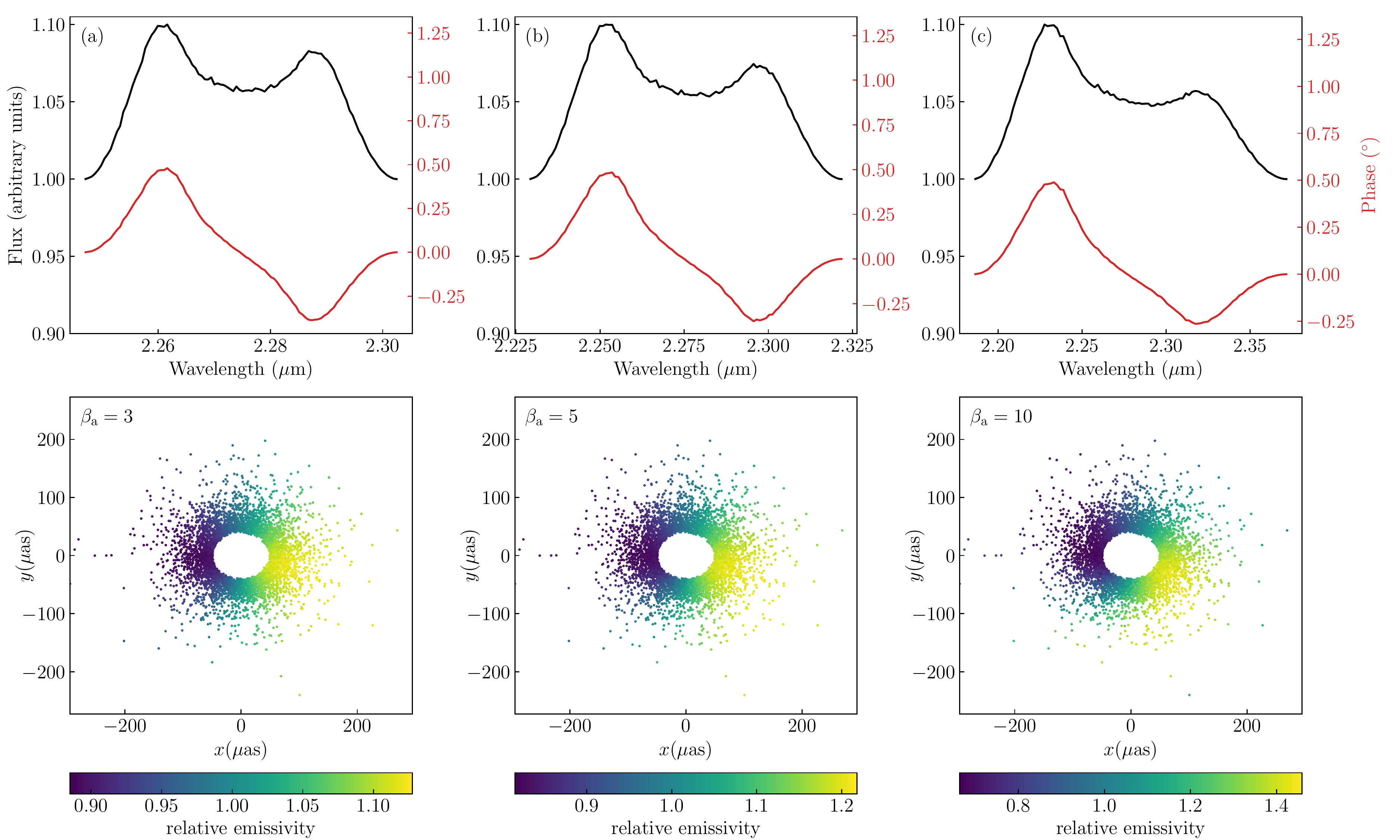}
    \caption{Dependence of line profile and differential phase curve on the rotational velocity of the binary black hole. \label{fig:para_beta}}
\end{figure}

Next, we adjust the rotational velocity of the binary from $0.05$ to $0.03$ and $0.10$, and the effect is presented in Fig. \ref{fig:para_beta}.
A larger rotational velocity will lead to a more significant boosting effect, and so a more asymmetric profile and phase curve.
However, we should note that if the rotational velocity is large enough to generate a spiral structure in the surface brightness distribution, the asymmetry will weaken again.

\begin{figure}
    \centering
    \includegraphics[width=\textwidth]{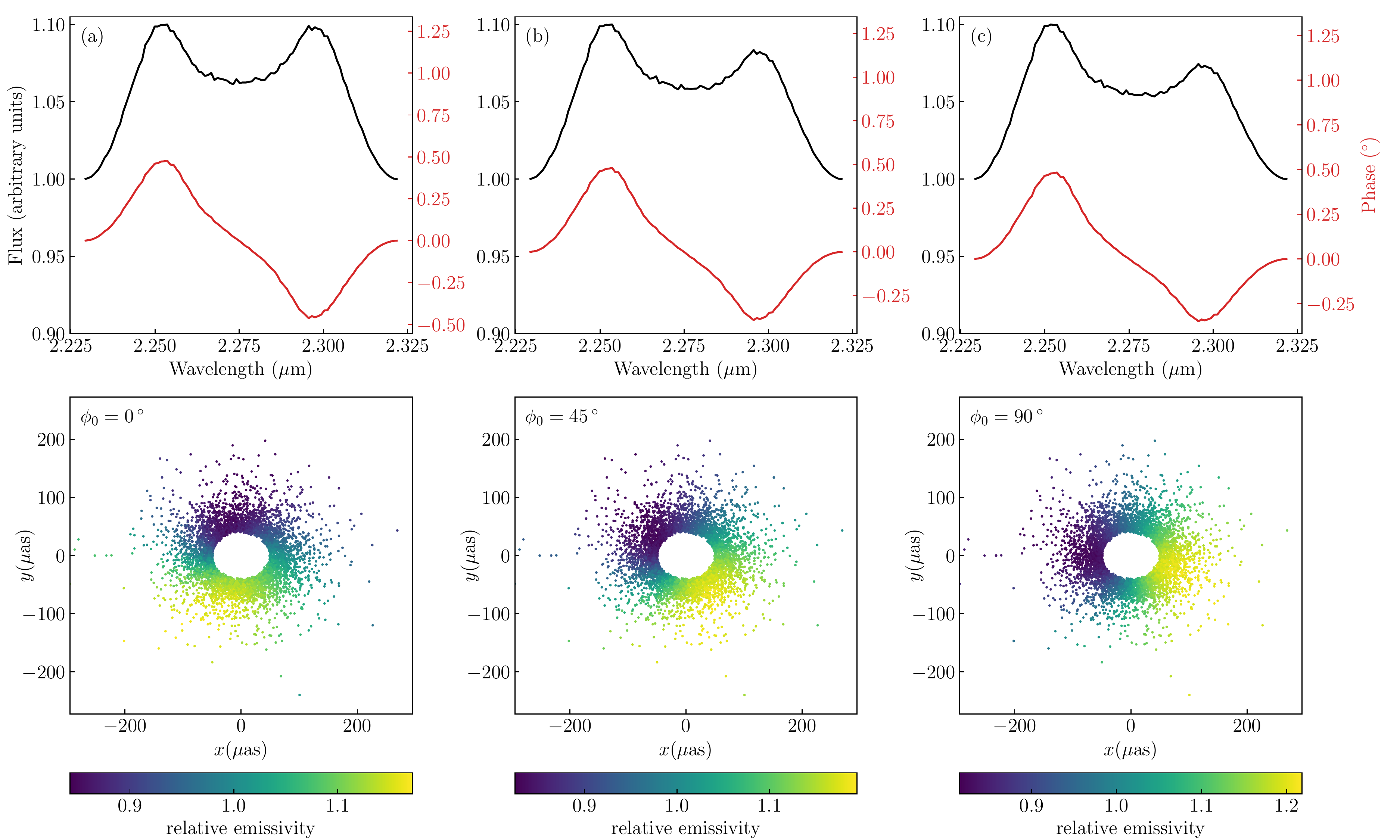}
    \caption{Dependence of line profile and differential phase curve on the orbital phase. \label{fig:para_phi}}
\end{figure}

The effect of the orbital phase of the binary on the profile and phase curve is shown in Fig. \ref{fig:para_phi}.
When the direction of the LOS velocity gradient is perpendicular to that of the emissivity gradient, as illustrated in Fig. \ref{fig:para_phi}(a), the profile and phase curve will be symmetric. 
As the binary rotates, the emissivity gradient will also change its direction accordingly, while the velocity gradient is always parallel to the long axis of the BLR's projection.
The profile and phase curve will reach the maximum asymmetry when the gradient of emissivity is parallel with that of LOS velocity.
So the profile and phase curve vary over the same period as the orbital period.

\begin{figure}
    \centering
    \includegraphics[width=\textwidth]{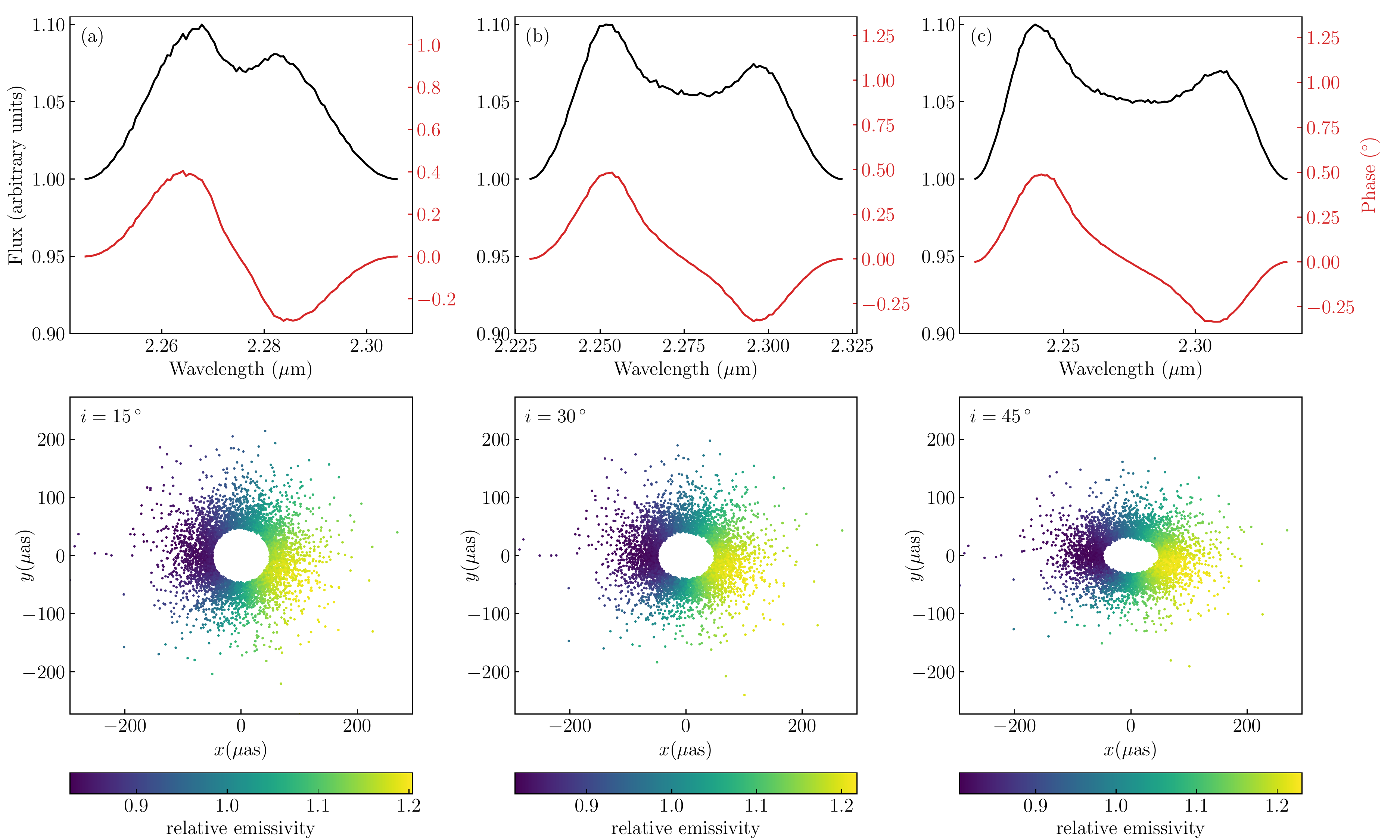}
    \caption{Dependence of line profile and differential phase curve on the inclination angle. \label{fig:para_inc}}
\end{figure}

When the inclination angle increased, the projected velocity of clouds will get larger and so broaden the line profile and phase curve, as shown in Fig. \ref{fig:para_inc}.
The doppler boosting effect mainly depends on the projected velocity of the binary ``seen'' by the BLR clouds, and so the asymmetry of the profile and phase curve is insensitive to the observer's inclination.

\begin{figure}
    \centering
    \includegraphics[width=\textwidth]{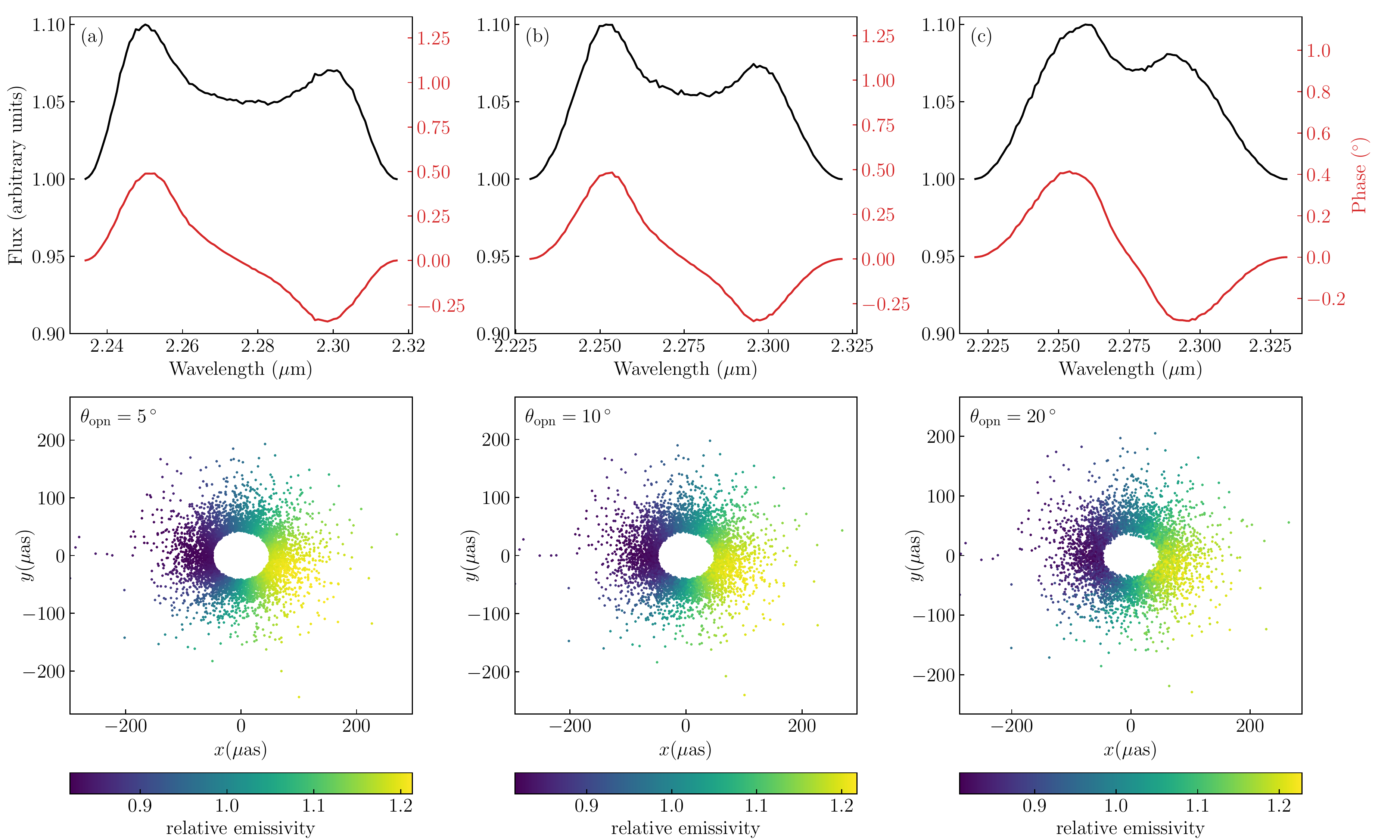}
    \caption{Dependence of line profile and differential phase curve on the opening angle of the BLR. \label{fig:para_opn}}
\end{figure}

Finally, we consider the effect of the opening angle of the BLR.
As the BLR gets thicker, the emissivities and LOS velocities of clouds will vary across the vertical direction of the BLR.
Subsequently, the inhomogeneity of emissivities and velocities will be smoothed out when projected onto the celestial sphere, leading to a more symmetric line profile and phase curve.
Meanwhile, double peaks in the line profile will also disappear for large opening angles, which is similar to the case of a single black hole.
We should emphasize that our model only applies to thin BLRs. 
When clouds deviate significantly from the orbital plane of the binary, the stabilities of their orbits need to be further clarified, which may lead to a more complex BLR model.

\subsection{Mock data analysis}
\begin{figure}
    \centering
    \includegraphics[width=\textwidth]{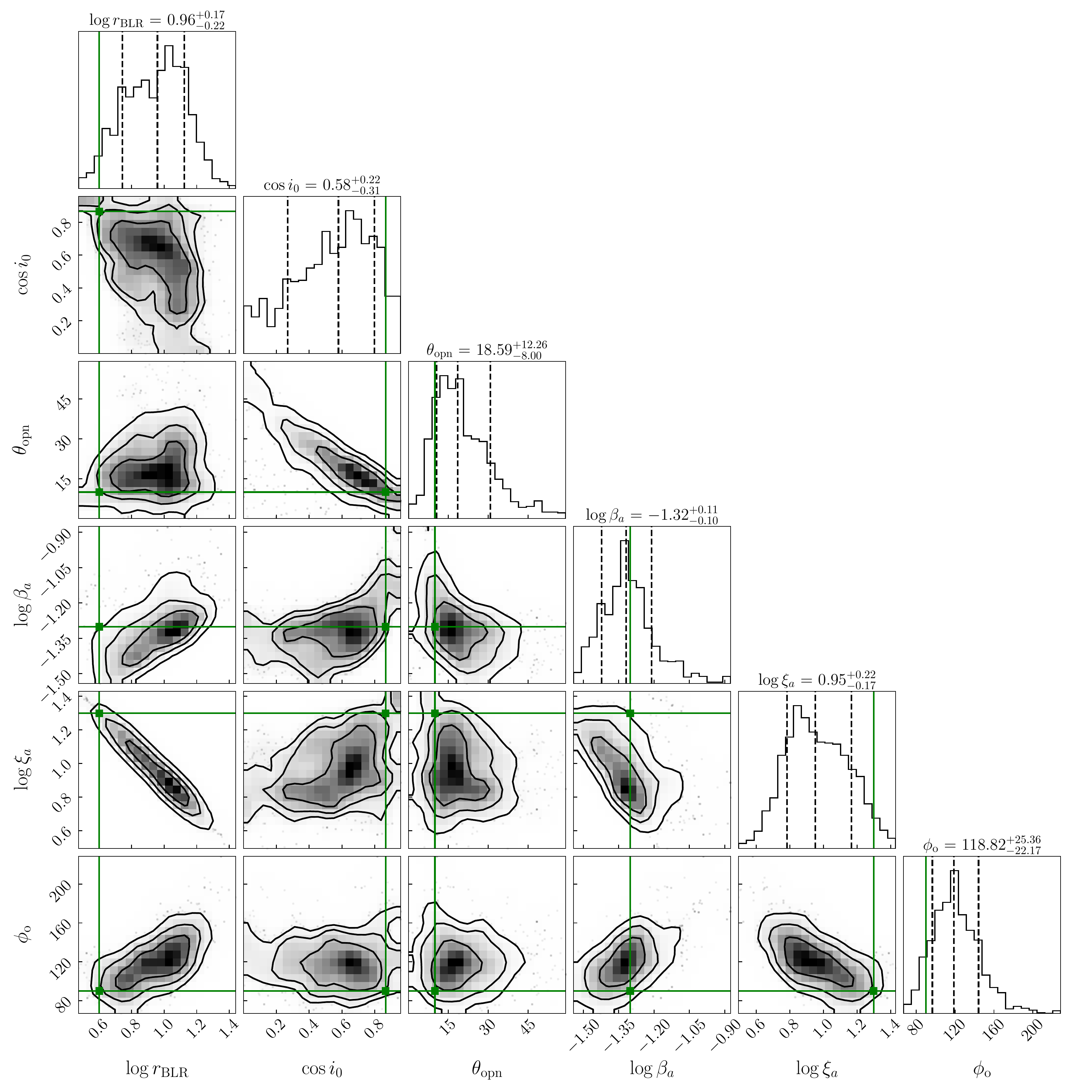}
    \caption{The posterior distribution of model parameters obtained by fitting the circumbinary BLR model to the mock SA data. 
    The median values with error bars at $1\sigma$ level of all parameters are given on the tops of panels. 
    Contours in two-dimensional distribution are at $1\sigma$, $1.5\sigma$, and $2\sigma$, respectively.
    The green lines represent input values. \label{fig:post}}
\end{figure}

The dependence of line profiles and differential phase curves on model parameters of circumbinary BLR indicates the possibility to infer the orbital parameters of the CB-SMBH through SA.
To see that quantitatively, we use the fiducial values in Table \ref{tab:CB-BLR} to generate mock data for SA observations \footnote{We change the luminosity ratio to $0.1$ to avoid boundary effect.}.
The details of the mock data generation can be found in the Appendix.
Given model parameters $\{\vb*{\Theta}\}$, the likelihood function of data set $\mathscr{D}$ is 
\begin{equation}
    P(\mathscr{D}|\vb*{\Theta}) = \prod_{j=1}^{N_{\lambda}} \frac{1}{\sqrt{2\pi \sigma_j^2}} \exp{-\frac{[F_{\mathrm{data},j} - F_{\mathrm{model},j}(\vb*{\Theta}) ]^2}{2\sigma_j^2}} \times  \prod_{i=1}^{N_{\rm b}} \prod_{j=1}^{N_{\lambda}} \exp{-\frac{[\phi_{\mathrm{data},ij} - \phi_{\mathrm{model},ij}(\vb*{\Theta}) ]^2}{2\sigma_{ij}^2}},
\end{equation}
where $N_{\lambda}$ and $N_{\rm b}$ are numbers of wavelength channels and interferometer baselines respectively, $F_{\mathrm{data},j}$ and $F_{\mathrm{model},j}(\vb*{\Theta})$ are input and predicted value of the flux at the $j$-th wavelength channel, $\phi_{\mathrm{data},ij}$ and $\phi_{\mathrm{model},ij}(\vb*{\Theta})$ are input and predicted value of the differential phase at the $j$-th wavelength channel of the $i$-th baseline, and $\sigma_i$ and $\sigma_{ij}$ are corresponding uncertainties of the line profile and differential phase curve.
The prior distribution of the model parameters $P(\vb*{\Theta})$ are listed in Table \ref{tab:CB-BLR}.
In light of Bayes's theorem, the posterior probability distribution for model parameters is
\begin{equation}
    P(\vb*{\Theta}|\mathscr{D}) \propto P(\vb*{\Theta})P(\mathscr{D}|\vb*{\Theta}),
\end{equation}
which can be sampled numerically using diffusive nested sampling (DNest) algorithm \citep{brewer2011}.
The result is shown in Fig. \ref{fig:post}.

For most model parameters, the true values are within or deviate somewhat from the $1\sigma$ interval of the posterior probability distribution.
The relative size of BLR $r_{\rm BLR}$ and the angular size of the binary separation $\xi_a$ are highly correlated. Only their product, the angular size of the BLR, can be determined quite accurately from the amplitude of the differential phase curve.
In our simulation, $r_{\rm BLR}$ is overestimated by by a factor of $2$ while $\xi_a$ is underestimated by about $\SI{50}{\percent}$.
The inclination of LOS $i_0$ and opening angle of BLR $\theta_{\rm opn}$ are also correlated since both decreasing $i_0$ and increasing $\theta_{\rm opn}$ can make the system more symmetric and weaken the double-peak structure in the profile.
The rotational velocity of the binary $\beta_a$ can be measured fairly accurately from the degree of asymmetry of the profile and phase curves.
Its correlation with the inclination angle is weak since clouds in the BLR always view the accretion disks edge-on, independent of the LOS of the observer.
However, the width of the profile is highly dependent on the inclination.
The severe uncertainty of the inclination makes it arduous to determine the relative size of the BLR accurately.

\section{Discussion}
\subsection{Parameter space and uncertainty of orbital parameters}
In this work, we mainly focus on the interferometric signatures of circumbinary BLRs around CB-SMBHs.
Binary separations are required to be large enough for nonuniformity in BLR to be resolved spatially, but also small enough such that Doppler boosting caused by orbital velocities is detectable, which will be of $0.01 - 0.1 \si{pc}$ generally.
To detect the GWs from the binary through PTA in the near future, we further require massive enough ($M_{\bullet} > 10^{8.5} M_{\sun}$) binaries with non-extreme mass ratios.
Circumbinary BLRs around CB-SMBHs may also be detected through line profiles \citep{nguyen2016} and RM \citep{ji2021}.
In both cases, we only require the orbital velocities to be large enough to be resolved spectroscopically or lead to a significant Dopper boosting effect, and so may explore a larger parameter space.
For binary with separation $\gtrsim 0.1 \si{pc}$, the BLR of each individual black hole will be detached.
If the mass ratio is not extreme and each of the binary BLRs has similar radiation flux, we can identify them through RM \citep{wang2018} and SA \citep{songsheng2019}.
While for binaries with extreme mass ratio, they can be selected from large populations of AGNs spectroscopically through time-dependent velocity offsets in line profiles \citep{kelley2021}.

\begin{figure}
    \centering
    \includegraphics[width=\textwidth]{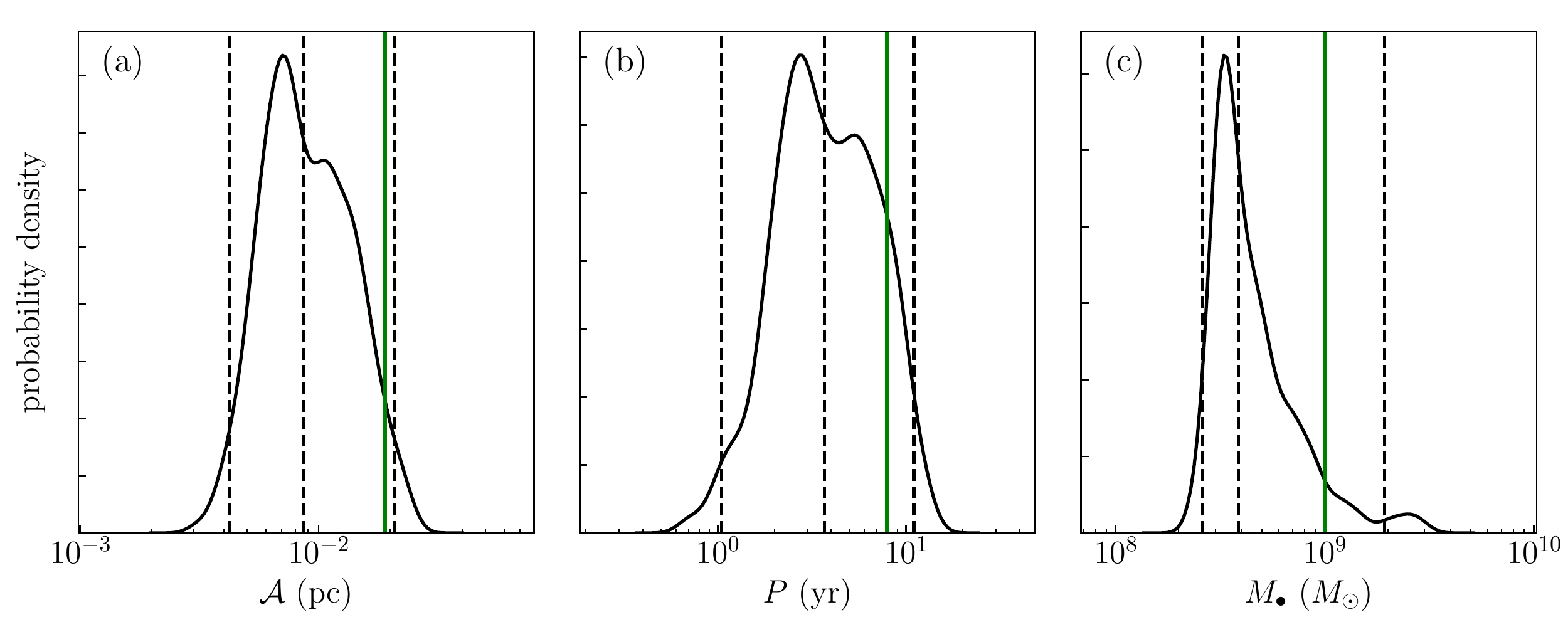}
    \caption{The posterior distribution of binary separation, orbital period, and total mass converted from distribution of model parameters. 
    The dashed lines are the 2\%, 50\%, and 98\% quantiles, while the green line is the true value. \label{fig:orbits}}
\end{figure}

The orbital parameters of the binary, such as binary separation and total mass, can be recovered from the posterior distribution of model parameters.
% using Eq. (\ref{eq:binary_separation}) - (\ref{eq:binary_mass})
The result is shown in Fig. \ref{fig:orbits}.
Input binary separation, orbital period, and total mass are all within the $2\sigma$ confidence interval of the posterior distribution, but they are underestimated by about $50\%$ systematically.

The major origin of the bias is the degeneracy between the relative BLR size and binary separation.
With SA observation, only their product, the absolute size of the BLR, can be determined accurately.
Extra information or data set is needed to break their degeneracy.
RM provides a supplementary perspective to probe the structure of the BLR along the LOS.
It has been demonstrated that joint analysis of RM and SA data can improve the measurement accuracy of black hole mass in the case of a single black hole \citep{wang2020}.
We would apply the joint analysis to check whether the uncertainty of the binary separation can be improved in a separate paper.

The minor origin of the bias is the large uncertainty of the inclination angle, which is always hard to measure accurately using line profiles.
The uncertainty of the inclination also contributes to the scatter of virial factor in RM \citep{collin2006}, hindering our accuracy in measuring masses of black holes through single-epoch spectrometry.
For AGNs with large-scale radio jets, the inclination can be constrained by superluminal motion \citep[e.g.][]{jorstad2017}.
However, the orbital motion of the binary may make jets from the two black holes wind with each other and form X-type lobes, leading to a more complicated scenario \citep[e.g.][]{merritt2002,cheung2007}.
Another way to measure the inclination angle is through spectropolarimetry.
Observations show that there are equatorial scattering regions in Type I AGNs \citep{smith2005}.
Unpolarized photons from the BLR are scattered in the direction of the LOS by electrons in the scattering regions and gain certain degrees of polarization.
As a result, the width of the polarized line profile is determined by the LOS velocities viewed by equatorial scattering regions and independent of the inclination of the observer's LOS \citep{baldi2016,songsheng2018}.
By comparing the width of total and polarized line profiles, the inclination angle can be constrained properly.

Another way to increase the measurement accuracy of orbital parameters is to use the periodicity of the binary. 
By monitoring the continuum variation modulated by the Doppler boosting effect, the orbital period can be determined accurately.
Fixing the period when fitting the BLR model to the SA data, the uncertainties of orbital parameters will decrease.
A more efficient method to constrain the orbital period is to conduct two SA observations several years apart and detect the difference between orbital phases.
Moreover, if the binary is close or massive enough, the GW signals generated by them could stand above the stochastic background and be resolved individually by PTA \citep{sesana2009}.
In this circumstance, the data of timing residuals and SA can be analyzed jointly to achieve multimessenger observations.
% Since the orbital period can be measured accurately using PTA, the combination of them will break the degeneracy between parameters and constrain the binary orbits reliably.

\subsection{Alternative model}
When the mean size of BLR is only a few times the separation between the binary, the region enhanced by the Doppler boosting is like a ``hot spot'' in the BLR.
A similar feature has also been reported in the RM observation of Arp 151 and can be explained by a BLR with warped-disk geometry \citep{bentz2010}.
The warp structure can expose or shield gas according to its position, causing the required azimuthal structure-enhancing responses.
It is hard to distinguish the circumbinary model from the warped-disk model by direct imaging of the BLR, let alone by a single epoch SA observation.
Besides detecting the periodicity of the light curve through the long-term monitoring campaign, it is possible to tell the two models apart using RM.
In the circumbinary BLR model, the source of ionizing radiation is in orbital motion rather than at rest in the center. 
The light travel time from the inner accretion disk to the BLR varies with azimuthal angle and time, displaying special features in the responses of broad emission lines to the continuum \citep{ji2021}.

If the size of BLR exceeds $\sim 10$ times the separation between the binary, spiral structures develop in the distribution of emissivities of BLR clouds.
Similar structures can also be generated by density waves of ionized gas in BLRs if the self-gravitating effects are important \citep{wang2022}.
In this case, not only the distribution of emissivities but also velocity gradients are modified by the ``spiral arms'', which is slightly different from the effect of Doppler boosting.
It is possible to distinguish them if the kinematics of the BLR gas is well constrained using SA or RM observations with high fidelity.

\subsection{Further improvement of the model}
In our formulation of the Doppler boosting effect, the intrinsic radiation flux variation of the accretion disk is neglected for simplicity.
Actually, the radiation from the $i$-th accretion disk $F_{{\rm AD}_i}$ in Eq. (\ref{eq:clouds_flux}) depends on the departure time $t_{ij}$ of the photon ionizing the $j$-th clouds.
For a rotating BLR, the difference of the departure time $t_{ij}$ for blueshifted and redshifted clouds is about $\mathcal{A}/c$.
Assuming the intrinsic variation can be described by a damped random walk, the characteristic amplitude of the variation on a timescale of $\mathcal{A}/c$ should be $\sigma \sqrt{\mathcal{A}/c}$, where $\sigma^2$ represents the variance in the light curve on short timescales \citep{kelly2009}.
If the mass of the black hole is about $10^9 M_{\sun}$, $\sigma^2$ will be of the order $\SI{e-4}{mag^2\,day^{-1}}$.
Taking $\mathcal{A} = \SI{e-2}{pc}$, we have $\Delta M_{\rm int} \sim \SI{5e-2}{mag} (\Delta F_{\rm int} / F \sim \SI{5}{\percent})$, which is much smaller than the anisotropy caused by the Doppler boosting effect.
However, the scatter of $\sigma^2$ is large among SMBHs with similar masses.
If the effect of the intrinsic variation is comparable to that of the Doppler boosting, we should include it in the model, and fit the SA data and continuum light curve simultaneously.

Secondly, the geometry of the circumbinary BLR is simply a circular disk in our model.
Nevertheless, the radiation of the accretion disk is highly anisotropic, so the ionization parameter varies with azimuthal angles.
In the direction where the radiation is enhanced, the distances of line-emitting clouds will be large, and vice versa.
As a result, the width of the line profile is also asymmetric in the red wing and blue wing, and the asymmetry of the differential phase curve is also enhanced.
To address the effect, the geometry of the BLR can be parameterized by an egg shape \citep[e.g.][]{narushin2021} rather than a circular.

Finally, we suppose that the dynamics of the BLR are stable and can be approximated by a Keplerian rotating disk.
Small deviations from this can be treated by adding random velocities to clouds, parameterized by ``turbulent'' velocities $\sigma_{\rm turb}$.
If the timescale for the BLR to reach the equilibrium state is longer than that of the orbital decay, we must include the evolution of the binary orbit in the dynamical simulation of cloud orbits to find a consistent velocity distribution for BLR clouds.
Furthermore, clouds in the BLR may be continuously replenished by torus \citep{wang2017} or accretion disk \citep{goad2012}, and so there could be clouds in dynamically unstable regions.
A more physical model is needed to connect the torus, BLR, and accretion disk and give a complete description of the gas environment around the SMBHs.

\section{Conclusion}

In this work, we establish a parameterized model for circumbinary BLRs, where the spatial and velocity distribution of clouds can be calculated conveniently and reliably.
Due to the Doppler boosting effect of accretion disks around individual black holes and finite travel time of light, structures like ``hot spot'' or ``spiral arm'' will appear in the distribution of emissivities, leading to asymmetric line profiles and differential phase curves.
Shapes of emission lines and phase curves vary with orbital parameters of the binary and geometries of the BLR.
As a result, orbital parameters can be recovered from mock SA data by fitting the model to the data, demonstrating its ability to identify CB-SMBHs and constrain their orbits in the future.
Uncertainties of parameters are mostly contributed by correlations between parameters.
Through joint analysis of multiple SA observations, RMs and PTAs, correlations can be broken to realize robust measurements of CB-SMBHs.

\begin{acknowledgments}
We are grateful to the members of the IHEP AGN group for enlightening discussions.
JMW thanks the support of the National Key R\&D Program of China through grant -2016YFA0400701, by NSFC through grants NSFC-11991050, -11991054, -11833008, -11690024, and by grant No. QYZDJ-SSW-SLH007 and No.XDB23010400. 
\end{acknowledgments}

\appendix
\section{Generation and fitting of the mock data}
\begin{figure}
    \centering
    \includegraphics[width=0.5\textwidth]{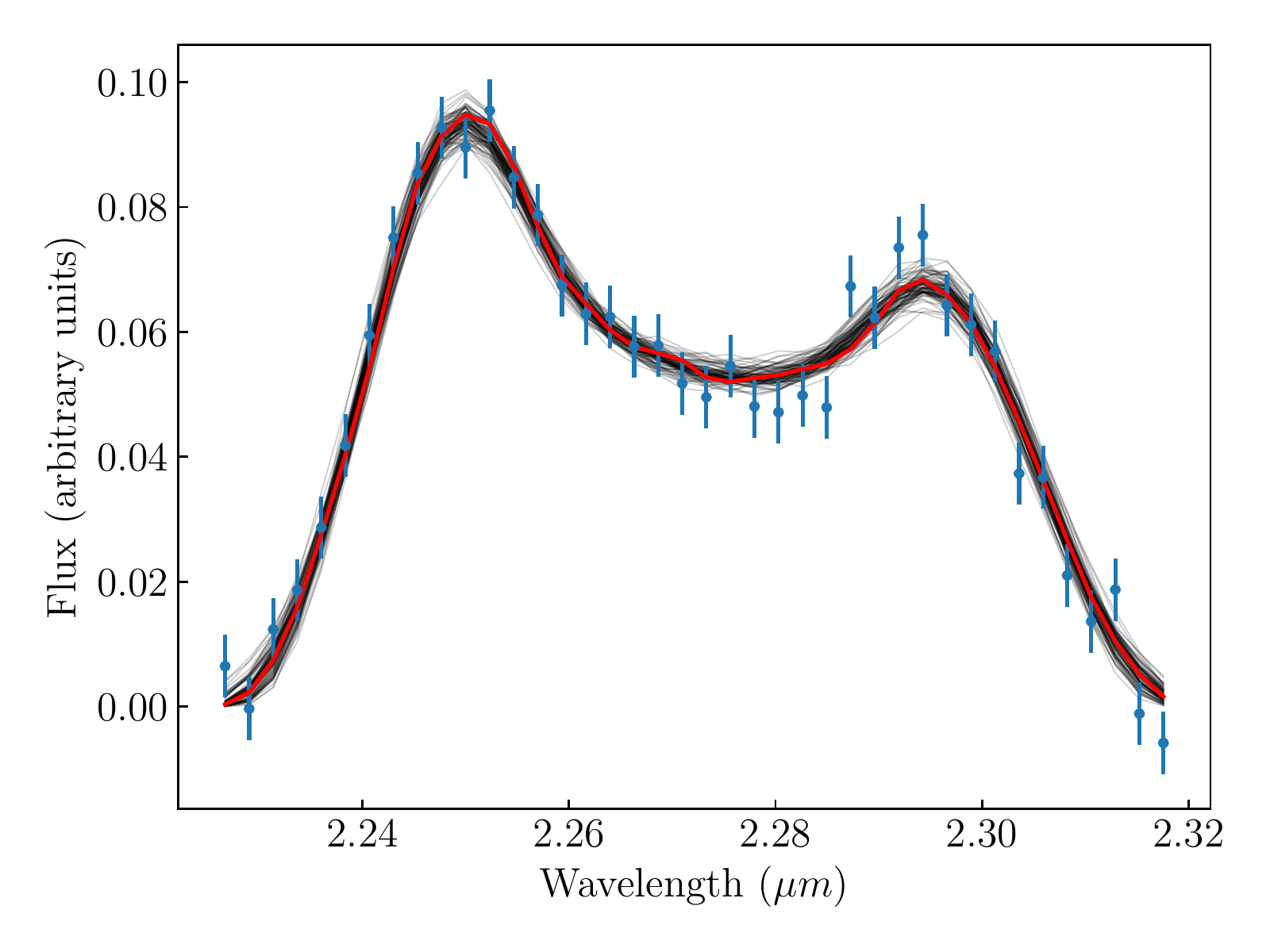}
    \caption{Mock line profile and fitting results. 
    The blue dots with error bars are data points with $1\sigma$ uncertainties.
    The thick red line corresponds to the best-fit curve, while the thin gray lines are fittings using model parameters randomly drawn from the posterior sample.
    \label{fig:mock_profile}}
\end{figure}

The line profile and differential phase curves for mock data analysis are calculated using the fiducial values of model parameters listed in Table \ref{tab:CB-BLR}.
The angular distance of the AGN is $\SI{200}{Mpc}$, and the corresponding redshift is $0.05$.
So the emission line used for SA observation in the K band is Br$\gamma$.
We normalize the profile of the Br$\gamma$ line so that its flux relative to the continuum is about $0.1$ at the peak.
The profile is then convolved using a Gaussian with FWHM of $\SI{4}{\mu m}$ to account for the effect of instrument broadening, which corresponds to the middle spectral resolution mode of GRAVITY.
The profile is sampled in $40$ wavelength bins between $\SI{2.22}{\mu m}$ and $\SI{2.32}{\mu m}$.
At each bin, the uncertainty of flux measurement is $\num{5e-3}$ relative to the continuum flux, which is also the typical value of line profiles obtained by GRAVITY.
The result is shown in Fig. \ref{fig:mock_profile}.

\begin{figure}
    \centering
    \includegraphics[width=\textwidth]{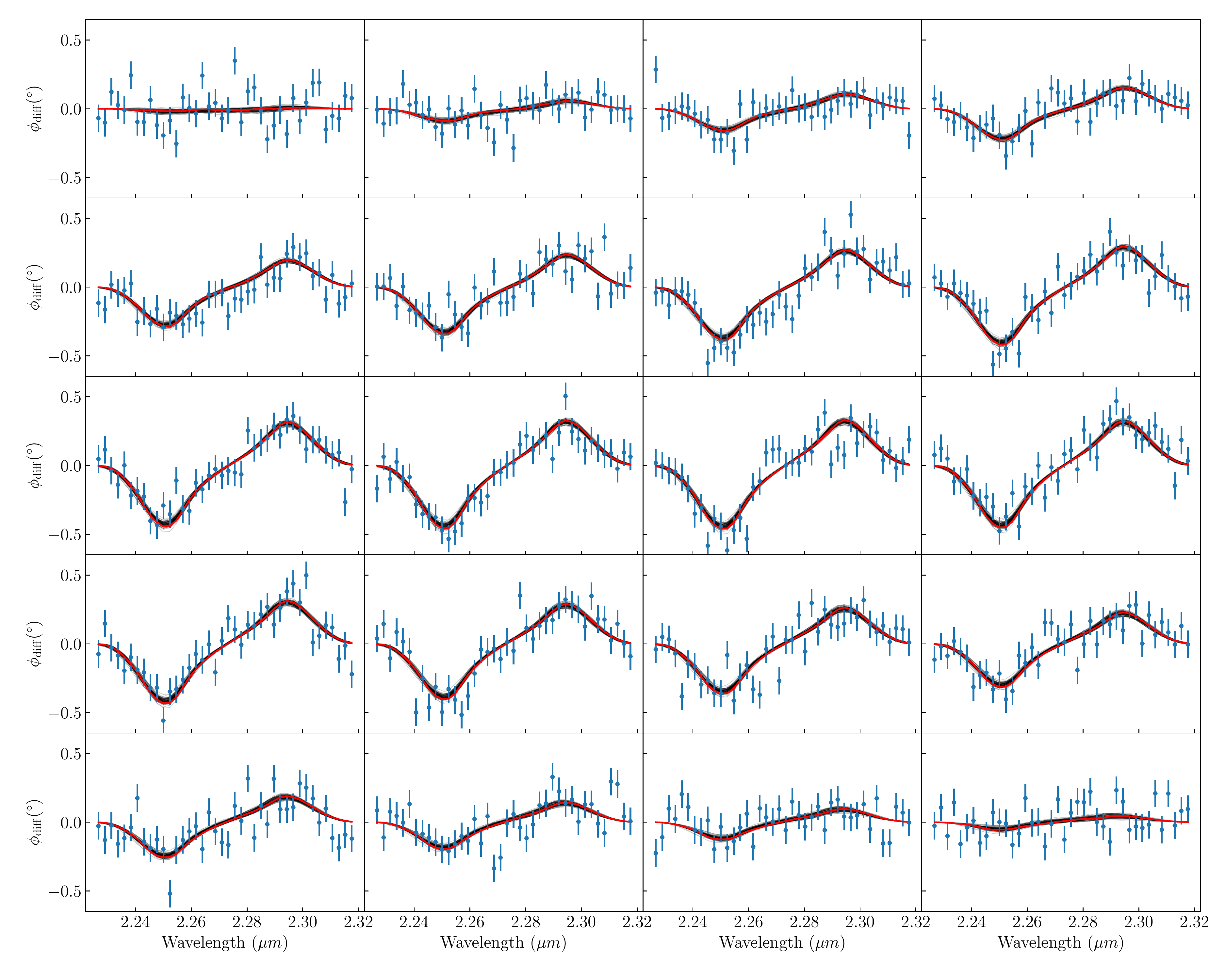}
    \caption{Mock differential phase curves and fitting results in each baseline.
    The blue dots with error bars are data points with $1\sigma$ uncertainties.
    The thick red line corresponds to the best-fit curve, while the thin gray lines are fittings using model parameters randomly drawn from the posterior sample. \label{fig:mock_phase}}
\end{figure}

To get differential phase curves, we need to project the displacement of the photocenter onto baselines in the $u-v$ plane, as indicated by Eq. \ref{eq:phase_curve}.
For simplicity, we assume there are $20$ baselines with length of $\SI{100}{m}$, and their azimuthal angles in the $u-v$ plane vary from $\SI{-90}{\degree}$ to $\SI{81}{\degree}$ uniformly.
Wavelength bins of differential phase curves are the same as those of the line profile.
At each bin, the uncertainty of interferometric phase measurement is $\SI{0.1}{\degree}$. 
The result is shown in Fig. \ref{fig:mock_phase}.

Once the mock data is available, we can sample the posterior probability distribution of model parameters in the Bayesian framework by fitting the model to the mock data using DNest.
We select 100 groups of parameters from the posterior sample randomly and calculate corresponding line profiles and differential phase curves, as shown by gray lines in Fig. \ref{fig:mock_profile} and \ref{fig:mock_phase}.
We also select the parameters with minimum chi-square from the posterior sample, and the best-fit curves predicted by them are shown as red lines in Fig. \ref{fig:mock_profile} and \ref{fig:mock_phase}.

\section{Sampling the posterior distributions of model parameters}
\begin{figure}
    \centering
    \includegraphics[width=0.83\textwidth]{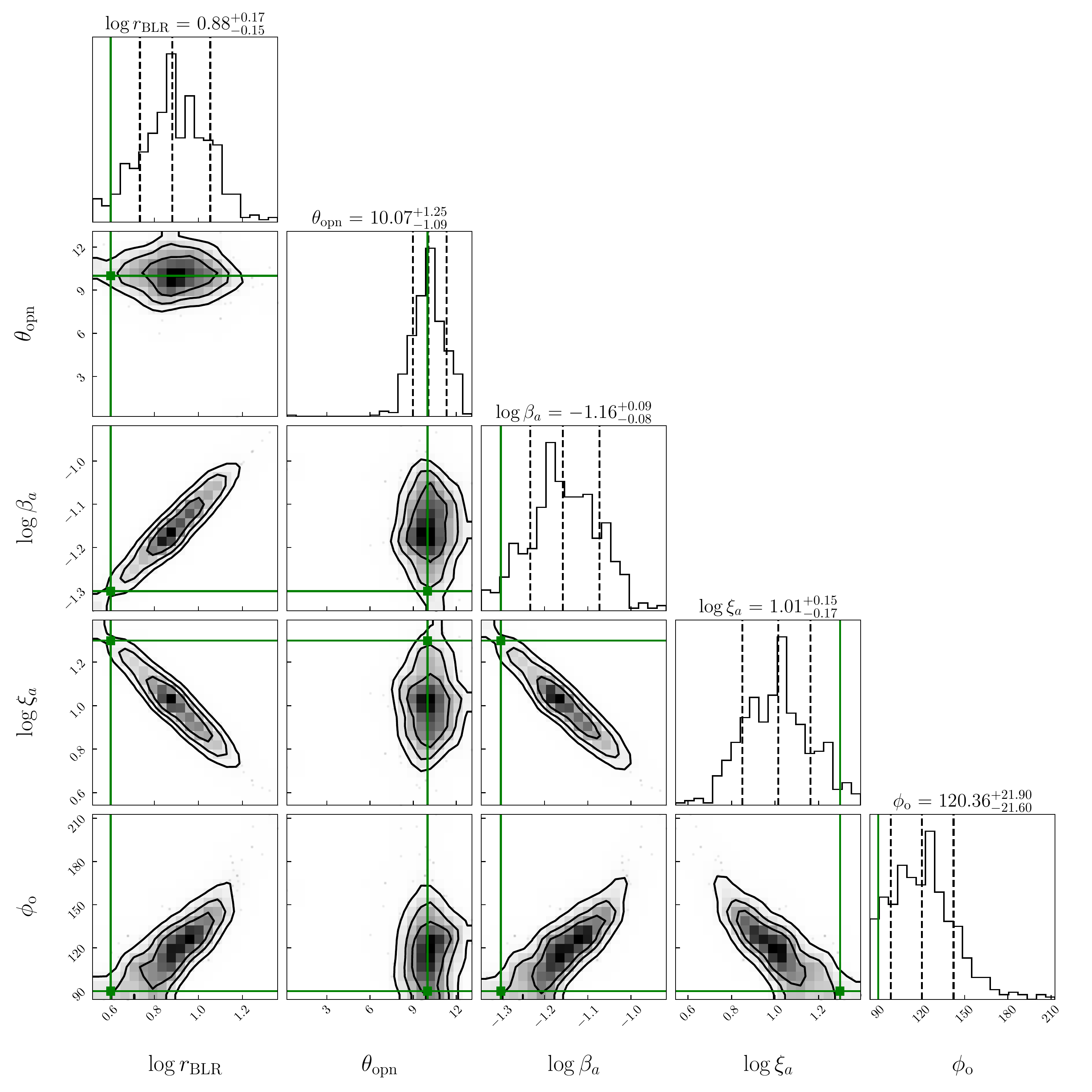}
    \caption{The posterior distribution of model parameters with a fixed inclination. 
    \label{fig:post_1}}
\end{figure}

The degeneracies between model parameters are the potential cause of discrepancies between the input and reconstructed values of model parameters when sampling the posterior distributions.
As shown in Fig. \ref{fig:post}, the $i_0 - \theta_{\rm opn}$ degeneracy and the $r_{\rm BLR} - \xi_a$ degeneracy are both significant.
Firstly, we fix the inclination to the input value and then sample the probability distribution of the remaining parameters.
The result is shown in Fig. \ref{fig:post_1}.
Just as we expect, the reconstructed value of the opening angle is exactly the same as the input one.
However, although the reconstructed values of the BLR size and binary separation are closer to the input ones than those in Fig. \ref{fig:post}, discrepancies between the input and reconstructed values are still larger than the $1\sigma$ uncertainties of the probability distributions.
So the degeneracy between the inclination and opening angle is not the primary cause of the biases in sampling the posterior distributions.

\begin{figure}
    \centering
    \includegraphics[width=0.83\textwidth]{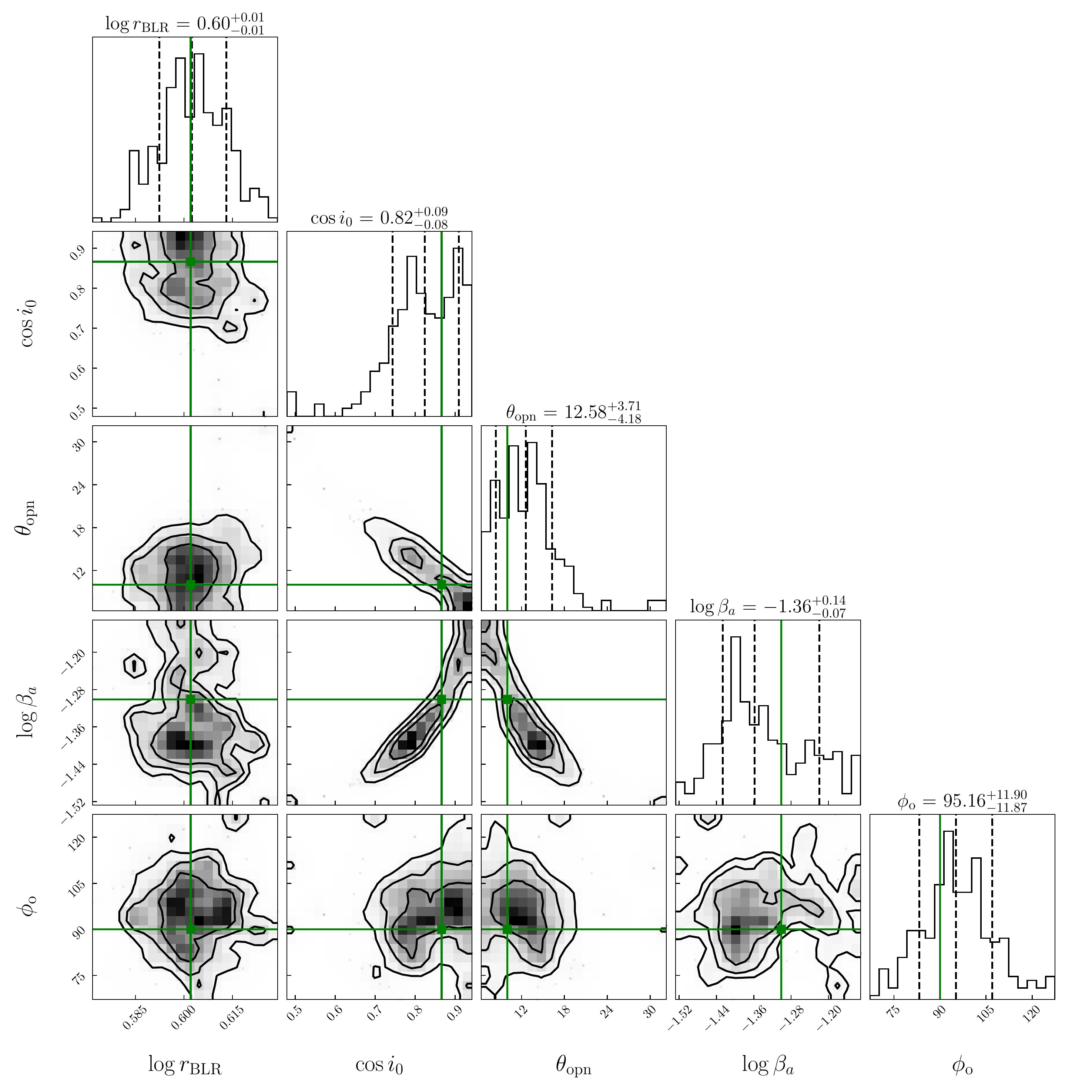}
    \caption{The posterior distribution of model parameters with a fixed binary separation. 
    \label{fig:post_2}}
\end{figure}

Next, we fix the binary separation to the input value and conduct the sampling.
The result is shown in Fig. \ref{fig:post_2}.
The input value of the BLR size coincides with the center of the posterior distribution, and the differences between reconstructed and input values for the inclination and the opening angle are also within the $1\sigma$ uncertainties of the probability distributions.
We can conclude that the biases are mainly due to the degeneracy between the binary separation and the BLR size.

\begin{figure}
    \centering
    \includegraphics[width=\textwidth]{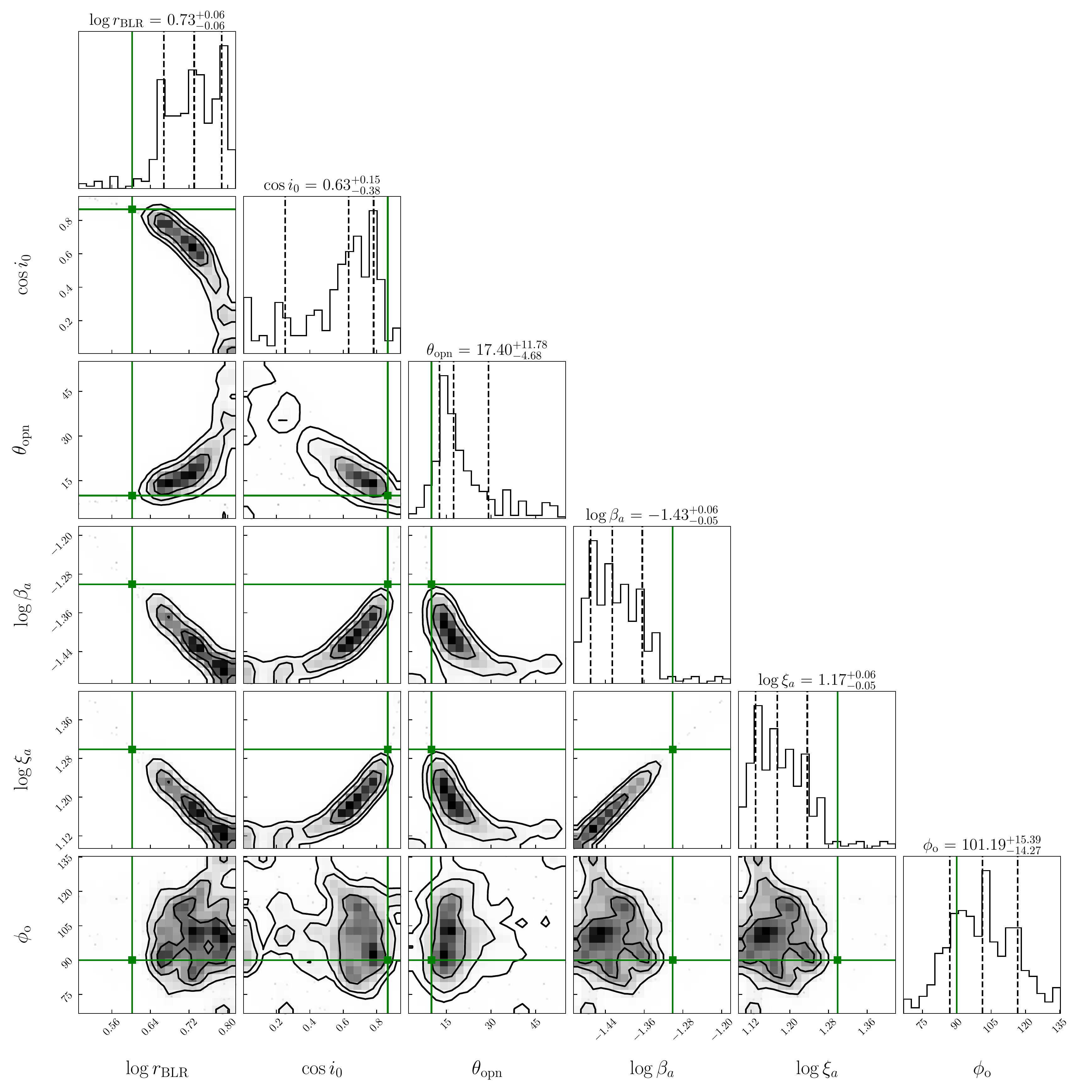}
    \caption{The posterior distribution of model parameters with a fixed binary period. 
    \label{fig:post_3}}
\end{figure}

We further fix the period of the binary to the input value and rerun the program.
The result is shown in Fig. \ref{fig:post_3}.
The $i_0 - \theta_{\rm opn}$ degeneracy and the $r_{\rm BLR} - \xi_a$ degeneracy still exist and biases are larger than the $1\sigma$ uncertainties of the posterior distributions.
But the uncertainties of posterior distributions of the BLR size and binary separation are improved by a factor of $3$ and so the estimation becomes more accurate.

\begin{figure}
    \centering
    \includegraphics[width=\textwidth]{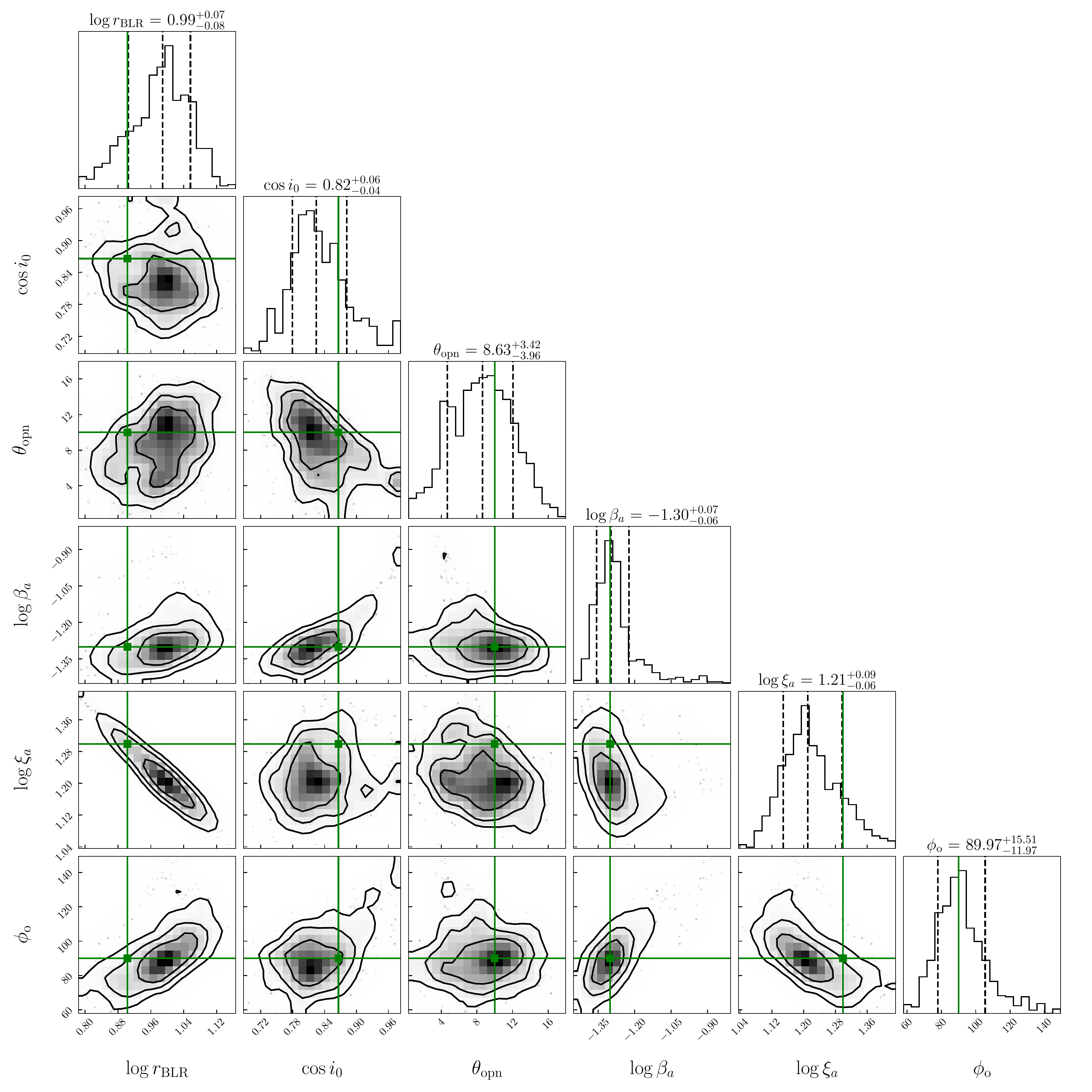}
    \caption{The posterior distribution of model parameters with larger BLR size. 
    \label{fig:post_4}}
\end{figure}

We now change the relative size of the BLR from the fiducial value of $r_{\rm BLR} = 4$ to $8$, simulate the new line profile and differential phase curve, and rerun the sampling process.
The result is shown in Fig. \ref{fig:post_4}.
The BLR size and binary separation are over-estimated and under-estimated by about $1\sigma$ respectively.
But the uncertainties of their posterior distributions are smaller than those in Fig. \ref{fig:post}.
Meanwhile, the reconstructed values of the inclination and the opening angle also match the input values within $1\sigma$ uncertainties.
So for larger BLR sizes, biases caused by the degeneracy between the binary separation and the BLR size are partially relieved, maybe because the amplitude of the phase curve gets larger.

\begin{figure}
    \centering
    \includegraphics[width=\textwidth]{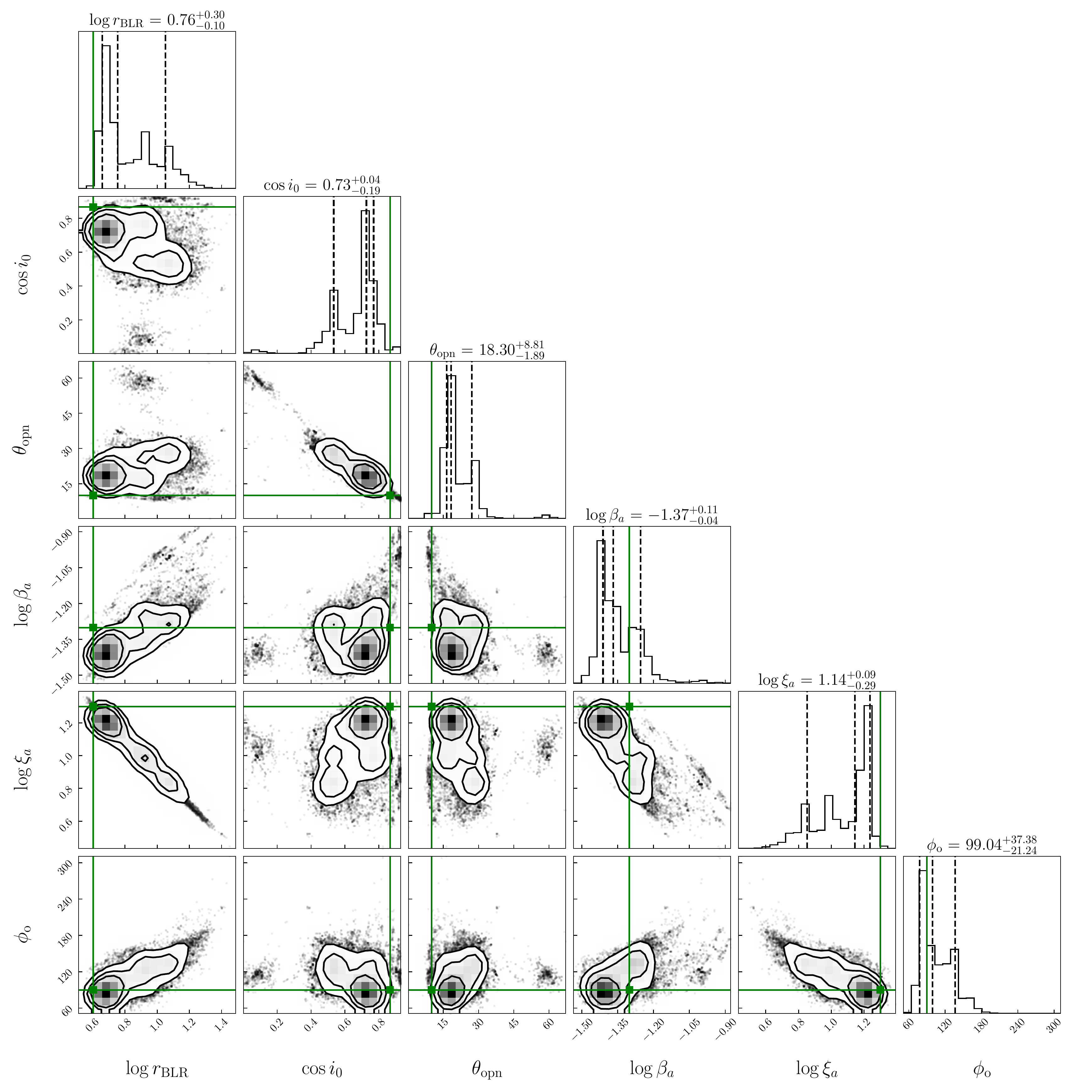}
    \caption{The posterior distribution of model parameters obtained by dynamical nested sampling method. 
    \label{fig:post_5}}
\end{figure}

Finally, we apply another sampling method, dynamical nested sampling method\citep{higson2019,speagle2020}, to construct the posterior distributions of model parameters for comparison.
The result is shown in Fig. \ref{fig:post_5}.
Similarly, the reconstructed values of most parameters differ from input values by more than one sigma.
However, the posterior distributions of the BLR size,  the inclination, and the binary separation show prominent multiple peaks, and the major peaks match the input values quite well.
The result demonstrates the fitting of the model to the data has degenerate solutions, which cause biases when inferring the model parameters.

\newcommand{\noop}[1]{}
\bibliography{ms}{}
\bibliographystyle{aasjournal}



\end{document}